\documentclass[amsmath,amssymb,aps,prd,preprint,superscriptaddress,showpacs]{revtex4-1}
\usepackage[dvipdfmx]{graphicx}
\usepackage{multirow}
\usepackage{times}
\usepackage[dvipdfmx]{color}
\allowdisplaybreaks[1]
\usepackage[normalem]{ulem}  % \sout{old text} for strikeout
\newcommand{\com}[1]{{\color[rgb]{0,0,1}{#1}}}

\renewcommand\sout{\bgroup \color{red} \ULdepth=-.5ex \ULset}
\newcommand{\bs}{\boldsymbol}
\begin{document}
\preprint{KUNS-2623}

\title{
Second-order hydrodynamics for fermionic cold atoms
\\
---Detailed analysis of transport coefficients and relaxation times---
}

\author{Yuta Kikuchi}
\email[]{kikuchi@ruby.scphys.kyoto-u.ac.jp}
%\thanks{}
\affiliation{Department of Physics, Faculty of Science, Kyoto University,
Kyoto 606-8502, Japan.}
%%%%%
\author{Kyosuke Tsumura}
\email[]{kyosuke.tsumura@fujifilm.com}
\affiliation{Analysis Technology Center,
  Research \& Development Management Headquarters,
  Fujifilm Corporation,
  Kanagawa 250-0193, Japan.}
%\thanks{}
%%%%%
\author{Teiji Kunihiro}
\email[]{kunihiro@ruby.scphys.kyoto-u.ac.jp}
\affiliation{Department of Physics, Faculty of Science, Kyoto University,
Kyoto 606-8502, Japan.}
%\thanks{}

\date{\today}

\begin{abstract}
We give a detailed derivation of the second-order (local) hydrodynamics for 
Boltzmann equation with an external force by using the renormalization group method. 
In this method, we solve the Boltzmann equation faithfully to extract the hydrodynamics without recourse to any ansatz.
Our method leads to microscopic expressions of not only all the transport coefficients 
that are of the same form as those in  Chapman-Enskog method but also
those of the viscous relaxation times $\tau_i$ that  admit physically natural interpretations. 
As an example, we apply our microscopic expressions to calculate the transport coefficients and the
relaxation times of the cold fermionic atoms in a quantitative way, where the transition probability 
in the collision term is given explicitly in terms of the $s$-wave scattering length $a_s$.
We thereby discuss the quantum statistical effects, temperature dependence, and scattering-length 
dependence of the first-order transport coefficients and the viscous relaxation times:
It is shown that as the temperature is lowered,
 the transport coefficients and the relaxation times
increase rapidly because Pauli principle acts effectively.
%tends to make particle scatterings other than 
%the forward one forbidden. 
On the other hand, as $a_s$ is increased, these quantities
 decrease and become vanishingly small at unitarity because of the strong coupling.
The numerical calculation  shows that the relation $\tau_\pi=\eta/P$, 
which is derived in the relaxation-time approximation and used in most of literature without almost any foundation, 
turns out to be satisfied quite well, while the similar relation for
the relaxation time $\tau_J$ of the heat conductivity is satisfied only approximately with a considerable error.
\end{abstract}

\maketitle

\section{
	Introduction
	}
\label{sec:sec1}

The hydrodynamic equation 
is expressed in terms of macroscopic quantities such as the pressure,
 particle number density, and fluid velocity,
 and takes a universal form irrespective of the microscopic dynamics of the system.
 The detailed microscopic properties of the system are renormalized into transport coefficients such as the shear viscosity, heat conductivity and so on. 
Therefore, the elaborate investigation of the transport coefficients
is one of the most important tasks to reveal the microscopic properties of the fluid. 
For instance, the fluid with a tiny shear viscosity is realized in the experiment of 
ultracold Fermi gases at the unitarity
 \cite{Ohara:2002observation,Kinast:2004,Bartenstein:2004,Schafer:2007,Cao:2010wa,Elliott:2014}: 
The value of its shear viscosity is close to a quantum bound that is theoretically proposed \cite{Policastro:2001yc,Kovtun:2004de}, 
implying the realization of the strongly correlated systems at the  unitarity
\cite{Gelman:2005,Bruun:2007etas,Rupak2007,Enss:2011,Hao:2011,Enss:2012}.
The hydrodynamic behavior with a small shear viscosity is also discovered 
in the ultra-relativistic heavy ion collision experiments 
at the Relativistic Heavy Ion Collider (RHIC) at the Brookhaven National Laboratory 
and the Large Hadron Collider (LHC) at CERN, which may again 
suggest that the created matter, i.e., Quark-gluon plasma (QGP) is a strongly coupled system (see Refs.~\cite{Schafer:2009,Adams:2012}, for instance).
It is noteworthy that, in spite of very large difference of the energy scale, 
these systems share common hydrodynamic properties, 
and the hydrodynamic equation provides us with a unified way to study their dynamics.

However, there is a problem in the application of the Navier-Stokes equation, besides the causality problem typical
 to the relativistic hydrodynamics: In finite systems, there are the central region where the density
 is large enough to apply the naive viscous hydrodynamic equation, and the peripheral region where the 
naive hydrodynamic description breaks down due to the small density. In the latter, since the system 
slowly approaches the  thermal equilibrium state due to the lack of enough collision rate,
 we need to take into account more microscopic dynamics. To this end, we should incorporate 
the relaxation process of dissipative currents, which is characterized by viscous relaxation 
times \cite{Massignan:2005,Bruun:2005,Bruun:2007etas,Bruun:2007,Braby:2011,Chao:2012}. 
The second-order hydrodynamic equation describes the mesoscopic dynamics including the relaxation 
of the dissipative currents, in addition to the ordinary hydrodynamic behavior described 
by the Navier-Stokes equation.
It should be emphasized that, though the importance of the second-order hydrodynamics
 has been recognized and many attempts has been done to derive it, its formulation is still controversial
 \cite{levermore1996moment,karlin1998dynamic,struchtrup2003regularization,gorban2004invariant,torrilhon2009special,torrilhon2010hyperbolic}.

In this paper, we derive the second-order hydrodynamic equation from 
the Boltzmann equation for non-relativistic systems by using the renormalization group (RG) method
 \cite{Chen:1994zza,Chen:1995ena,Kunihiro:1995zt,Kunihiro:1996rs,kunihiro1998dynamical,
Kunihiro:1997uy,Kunihiro:1998jp,Boyanovsky:1998aa,Ei:1999pk,Boyanovsky:1999cy,Hatta:2001ui,Boyanovsky:2003ui,Kunihiro:2005dd}. 
In the RG method, we faithfully solve the Boltzmann equation 
and extract the hydrodynamics as a low-energy effective dynamics of the kinetic theory.
It has been applied to derive the first- and second-order hydrodynamic equations 
for both relativistic and non-relativistic systems
 \cite{Hatta:2001ui,Tsumura:2007ji,Tsumura:2012gq,Tsumura:2012kp,Tsumura:2013uma,Tsumura:2015fxa}, 
and desirable properties have been already shown for the resultant equation such as causality, 
stability, positivity of the entropy production rate, 
and the Onsager's reciprocal relation without imposing any assumption a priori
\cite{Tsumura:2015fxa,Kikuchi:2015swa}. 
Moreover, the microscopic expressions obtained for the transport coefficients 
such as the shear viscosity, heat conductivity and  so on coincide with those derived in the
celebrated Chapman-Enskog method, while the novel  microscopic expressions
of the viscous relaxation times written in terms of the relaxation functions
allow physically natural interpretations as the relaxation times.
As an extension of Ref.~\cite{Tsumura:2013uma}, we take account of the effect of quantum statistics 
and external forces.
As an application, we use our microscopic expressions to calculate 
the shear viscosity, heat conductivity, 
and viscous relaxation times of the stress tensor and heat flow
 of the cold fermionic atoms in a quantitative way, where the transition amplitude
in the collision term is given explicitly in terms of the $s$-wave scattering length $a_s$:
The microscopic expressions given in forms of correlation functions are converted to
linear integral equations, which we solve numerically without recourse to any approximation.
On the basis of the numerical results, 
we discuss the quantum statistical effects, temperature dependence, and scattering-length 
dependence of the first-order transport coefficients and the viscous relaxation times:
It is shown that at low temperatures,
quantum statistics acts so effectively  that 
the transport coefficients and the relaxation times
increase rapidly because Pauli principle almost forbids particle scatterings other than the forward ones.
On the other hand, as $a_s$ is increased up to unitarity, these quantities
 decrease monotonically and become vanishingly small at unitarity because of the strong coupling.
We also examine how well the relation $\tau_\pi=\eta/P$ 
and its analog for the heat conductivity are satisfied, where $\tau_\pi$ 
is the viscous relaxation time of the stress tensor, 
$\eta$ the shear viscosity, and $P$ the pressure.
These relations are obtained from the Boltzmann equation with use of the relaxation-time 
approximation (RTA), which has been widely applied to a lot of studies of the kinetic theory
\cite{Bruun:2007,Braby:2011,Chao:2012}.
Although the RTA might happen to be valid for a system close to the local equilibrium, 
its quantitative reliability is unclear and has  been hardly checked even apart from 
the fact that the relaxation times should have  different values depending on the realization process:
We are only aware of \cite{Schafer:2014} in which the validity of the RTA is analytically examined up 
to some approximations.
We show that the relation $\tau_\pi=\eta/P$ holds quite well, while the similar relation for
the relaxation time $\tau_J$ of the heat conductivity is satisfied only approximately with a considerable error.
Here we should mention that a brief report of the present work is already given in Ref.~\cite{kikuchi:2015letter},
and the present paper give the detailes of not only the analytic but also numerical calculations.

This paper is organized as follows: 
In Sec.~\ref{sec:sec2}, we briefly summarize the properties of the Boltzmann equation. 
In Sec.~\ref{sec:sec3}, we derive the second-order hydrodynamic equation by using RG method, and show the resultant equations and microscopic expressions of the transport coefficients. 
In Sec.~\ref{sec:sec4}, we reduce the microscopic expressions of the transport coefficients and the viscous relaxation times for the numerical calculations.
In Sec.~\ref{sec:sec5}, we show the numerical results of the transport coefficients and the viscous relaxation times, 
and discuss the physical properties of the numerical results with temperature and
the scattering length being varied. Convergence of the numerical results are confirmed in the last part of this section.
In Sec.~\ref{sec:sec6}, we give the concluding remarks.
In Appendix.~\ref{sec:app1}, we give a useful formula which is used to solve the Boltzmann equation.
In Appendix.~\ref{sec:app2}, we present the detailed derivation of the relaxation equation.

%%%%%%%%%%%%%%%%%%%%%%%%%%%%%%%%%%%%%%%%%%%%%%%%%%%%%%%%%%%%%%%%%%%%%

%%%%%%%%%%%%%%%%%%%%%%%%%%%%%%%%%%%%%%%%%%%%%%%%%%%%%%%%%%%%%%%%%%%%%
%\setcounter{equation}{0}
\section{
  Boltzmann equation
}
\label{sec:sec2}

In this section, we give a brief review of the Boltzmann equation and its properties.
The Boltzmann equation, which describes the time-evolution of the one-body distribution
 function $f_p(t,\bs{x})$ in the phase space, takes the following form:
\begin{align}
 \label{eq:BE1}
 \left(\frac{\partial}{\partial t} + \bs{v}\cdot \bs{\nabla} + \bs{F}\cdot \bs{\nabla}_p\right)
 f_p(t,\bs{x})=C[f]_p(t,\bs{x}),
\end{align}
with $\bs{v}\equiv \bs{p}/(2m)$ and the collision integral given by
\begin{align}
 \label{eq:collision}
 C[f]_p(t,\boldsymbol{x})
 &=\frac{1}{2}\int_{p_1}\int_{p_2}\int_{p_3}\mathcal{W}(p,p_1|p_2,p_3) 
 \big(\bar{f}_p\bar{f}_{p_1}f_{p_2}f_{p_3}-f_p f_{p_1}\bar{f}_{p_2}\bar{f}_{p_3}\big).
\end{align}
Here, we have introduced the notations $\int_p\equiv \int\mathrm{d}^3p/(2\pi)^3$
 and $\bar{f}\equiv 1+af$. $a$ represents quantum statistics, i.e., $a=-1(+1)$ 
for fermion (boson) and $a=0$ for the classical Boltzmann gas.
$\bs{F}$ represents the external force which particles experience. 
 In this paper, we consider the force driven by the scalar potential,
 $\bs{F}=-\bs{\nabla}E_p$ with $E_p=|\bs{p}|^2/(2m)+V(\bs{x})$, where any inhomogeneous mean fields may be incorporated into 
the potential $V$. $\mathcal{W}$ is
 a transition matrix given by
\begin{align}
 \label{eq:transition}
 \mathcal{W} &= |\mathcal{M}|^2 (2\pi)^4\delta(E_p+E_{p_1}-E_{p_2}-E_{p_3})
 \delta^3(\bs{p}+\bs{p}_1-\bs{p}_2-\bs{p}_3),
\end{align}
with the scattering amplitude $\mathcal{M}$.
The transition matrix has the following symmetries
\begin{align}
 &\mathcal{W}(p,p_1|p_2,p_3) = \mathcal{W}(p_2,p_3|p,p_1) 
 = \mathcal{W}(p_1,p|p_3,p_2) = \mathcal{W}(p_3,p_2|p_1,p).
\end{align}
By using these symmetries one finds that  the following identity is 
satisfied for an arbitrary spacetime dependent vector $\Phi_p(x)$:
\begin{align}
 \label{eq:coll_inv}
 \int_p \Phi_p C[f]_p &= \frac{1}{8}\int_p\int_{p_1}\int_{p_2}\int_{p_3}\mathcal{W}(p,p_1|p_2,p_3) 
 \nonumber\\
 &\times\big(\Phi_p+\Phi_{p_1}-\Phi_{p_2}-\Phi_{p_3}\big)
 \big(\bar{f}_p\bar{f}_{p_1}f_{p_2}f_{p_3}-f_p f_{p_1}\bar{f}_{p_2}\bar{f}_{p_3}\big).
\end{align}
When $\Phi_p$ vanishes Eq.~\eqref{eq:coll_inv}, $\Phi_p$ is called a collision invariant,
and any linear combination of conserved quantities, i.e.,
the particle number, momenta, and energy, is found to be a collision invariant. 
Accordingly, $\Phi_p=\alpha+\bs{\beta}\cdot\bs{p}+\gamma E_p\equiv \Phi_p^{\rm inv}(x)$ 
with the space\com{-}time dependent coefficients $\alpha$, $\bs{\beta}$ and $\gamma$ 
is a collision invariant.

In the Boltzmann theory, the entropy density and current are defined by
\begin{align}
 \{s, \bs{J}_s\} \equiv -\int_p \{1,\bs{v}\}\left(f_p\ln f_p -\frac{\bar{f}_p\ln\bar{f}_p}{a}\right),
\end{align}
which satisfies
\begin{align}
 \label{eq:entropy}
 \frac{\partial s}{\partial t}+\bs{\nabla}\cdot\bs{J}_s
 &=-\int_p\ln\left(\frac{f_p}{\bar{f}_p}\right)
 \left(\frac{\partial f_p}{\partial t} + \bs{v}\cdot \bs{\nabla} f_p\right)
 =-\int_p\ln\left(\frac{f_p}{\bar{f}_p}\right)C[f]_p.
\end{align}
In the second equality, we have utilized the fact that the force term does not contribute as
\begin{align}
 \int_p\ln\left(\frac{f_p}{\bar{f}_p}\right)\bs{F}\cdot\bs{\nabla}_p f_p
 &=-\bs{F}\cdot \int_p f_p {\bs{\nabla}}_p\ln\left(\frac{f_p}{\bar{f}_p}\right)
 =-\bs{F}\cdot \int_p f_p \left(\frac{{\bs{\nabla}}_pf_p}{f_p}-\frac{a{\bs{\nabla}}_pf_p}{\bar{f}_p}\right)
 \nonumber\\
 &=-\bs{F}\cdot \int_p \frac{{\bs{\nabla}}_pf_p}{\bar{f}_p}
 = - \frac{1}{a}\bs{F}\cdot \int_p\bs{\bs{\nabla}}_p \ln\bar{f}_p 
 =0.
\end{align}
where $\bs{F}$ is assumed to be independent of momentum and we have neglected surface terms.
From Eq.~\eqref{eq:entropy}, one finds that entropy is conserved if $\ln(f_p/\bar{f}_p)$ is 
a collision invariant $\Phi_p^{\rm inv}(x)$, i.e., $f_p=1/[\mathrm{e}^{-\Phi_p^{\rm inv}(x)}-a]$, which is reduced to the form of a local equilibrium distribution function,
\begin{align}
 f^\mathrm{st}_p
 &=\frac{1}{\mathrm{e}^{(m|\bs{v}-\bs{u}|^2/2+V-\mu)/T}-a}
 =\frac{1}{\mathrm{e}^{(|\bs{\delta p}|^2/(2m)+V-\mu)/T}-a}.
\end{align}
Here we have introduced the relative momentum $\bs{\delta p}$. The relative momentum $\bs{\delta p}$ and relative velocity $\bs{\delta v}$ against the fluid velocity $\bs{u}$ are defined by $\bs{\delta p} \equiv m\bs{\delta v} \equiv m\bs{v}-m\bs{u}$.

%%%%%%%%%%%%%%%%%%%%%%%%%%%%%%%%%%%%%%%%%%%%%%%%%%%%%%%%%%%%%%%%%%%%%

%%%%%%%%%%%%%%%%%%%%%%%%%%%%%%%%%%%%%%%%%%%%%%%%%%%%%%%%%%%%%%%%%%%%%
%\setcounter{equation}{0}
\section{
  Derivation of hydrodynamics
}
\label{sec:sec3}

The RG method is a general framework 
to identify (fewer) slow variables and extract their dynamics 
from the original complicated dynamics \cite{Chen:1994zza,Kunihiro:1995zt,Kunihiro:1996rs,Ei:1999pk}.
In this section, we solve the Boltzmann equation \eqref{eq:BE1} to derive the second-order hydrodynamic equation 
by applying the RG method.
The key to extract mesoscopic dynamics in addition to the slowest macroscopic dynamics is how to identify the excited modes, 
which is realized by utilizing the doublet scheme \cite{Tsumura:2013uma,Tsumura:2015fxa} to which we refer  
the details of the RG method applied to Boltzmann equation to extract hydrodynamics, although  any mean field is
not included in the Boltzmann equation there.

\subsection{
Solving the Boltzmann equation
}

We perform the perturbative calculation assuming that the mean free path is much smaller than the spatial variation of the potential term. We introduce the bookkeeping parameter $\epsilon$ to the Boltzmann equation \eqref{eq:BE1},
\begin{align}
 \label{eq:BE2}
 \left(\frac{\partial}{\partial t} + \epsilon\bs{v}\cdot \bs{\nabla} + \epsilon\bs{F}\cdot \bs{\nabla}_p\right) f_p(t,\bs{x})=C[f]_p(t,\bs{x}).
\end{align}
The parameter $\epsilon$ is interpreted as the Knudsen number defined by (mean free path)$/$(macroscopic length scale). 
Note that the external force is also assumed to be as small as spatial inhomogeneity since it is caused by the spatial variation of the external potential.

We expand the distribution function in the following perturbation series with respect to $\epsilon$,
\begin{align}
 \tilde{f}(t;t_0)=\tilde{f}^{(0)}(t;t_0)+\epsilon\tilde{f}^{(1)}(t;t_0)
 +\epsilon^2\tilde{f}^{(2)}(t;t_0)+\mathcal{O}(\epsilon^3),
\end{align}
and solve the Boltzmann equation order by order under the initial condition,
 \begin{align}
 \tilde{f}(t=t_0)&=f(t=t_0)
 =f^{(0)}(t_0)+\epsilon f^{(1)}(t_0)+\epsilon^2 f^{(2)}(t_0)+\mathcal{O}(\epsilon^3),
\end{align}
which is taken to be an exact solution of the Boltzmann equation \eqref{eq:BE2}.

The zeroth order equation with respect to $\epsilon$ reads
\begin{align}
 \frac{\partial}{\partial t}\tilde{f}_p^{(0)}(t;t_0)
 =C[\tilde{f}^{(0)}]_p(t;t_0).
\end{align}
Since we are interested in a slow motion in the asymptotic regime, we seek for the solution of the following equation,
\begin{align}
 \frac{\partial}{\partial t}\tilde{f}_p^{(0)}(t;t_0) =0,
\end{align}
which is equivalent to
\begin{align}
C[\tilde{f}^{(0)}]_p(t;t_0)= 0.
\end{align}
This equation leads to the condition
\begin{align}
 \bar{f}_p\bar{f}_{p_1}f_{p_2}f_{p_3} = f_p f_{p_1}\bar{f}_{p_2}\bar{f}_{p_3}.
\end{align}
By taking logarithm of both sides of this equation we note that the equation requires 
that $\ln(f_p/\bar{f}_p)$ should become  a linear combination of $\{1,\bs{p},E\}$ the coefficients which
are constant in time $t$ but may depend on the space coordinate $\bs{x}$ and the initial time $t_0$. 
From the discussion in the last section,
 we see that the zeroth order solution takes the form of a local equilibrium distribution function,
\begin{align}
 &\tilde{f}_p^{(0)}(t;t_0) = f^\mathrm{eq}_p(t_0)
 =\left(\exp\left[\frac{(m/2)|\bs{v}-\bs{u}(t_0,\bs{x})|^2-\mu_\mathrm{TF}(t_0,\bs{x})}{T(t_0,\bs{x})}\right]-a\right)^{-1},
\end{align}
where we have defined $\mu_{TF}(t_0,\bs{x})=\mu(t_0,\bs{x})-V(\bs{x})$.
Five integration constants $T(t_0,\bs{x})$, $\bs{u}(t_0,\bs{x})$, and $\mu(t_0,\bs{x})$ 
are would-be temperature, fluid velocity, and chemical potential, which 
will be lifted to dynamical variables eventually to characterize the slowest dynamics of resultant hydrodynamics.

Next, we consider the first order equation:
\begin{align}
 \frac{\partial}{\partial t} \tilde{f}^{(1)}(t)
 =f^{\mathrm{eq}}\bar{f}^{\mathrm{eq}}L
 (f^{\mathrm{eq}}\bar{f}^{\mathrm{eq}})^{-1}\tilde{f}^{\mathrm{(1)}}(t)
 +f^{\mathrm{eq}}\bar{f}^{\mathrm{eq}}F_0, 
\end{align}
with the initial condition
\begin{align}
 \tilde{f}^{(1)}(t=t_0,t_0)=f^{(1)}(t_0) \equiv f^\mathrm{eq}\bar{f}^\mathrm{eq}\Psi(t_0).
\end{align}
$\Psi(t_0)$ is to be determined later.
Here, the linearized collision operator $L$ and the inhomogeneous term $F_0$ are defined by
\begin{align}
 \label{eq:LCO}
 L_{pq}&\equiv
 (f^{\mathrm{eq}}_p\bar{f}^{\mathrm{eq}}_p)^{-1}
 \left.\frac{\delta}{\delta f_q}C[f]_p(t)\right|_{f=f^{\mathrm{eq}}}
 f^{\mathrm{eq}}_q\bar{f}^{\mathrm{eq}}_q
 \nonumber\\
 &=-\frac{1}{2\bar{f}^{\mathrm{eq}}_p}\int_{p_1}\int_{p_2}\int_{p_3}
 \mathcal{W}(p,p_1|p_2,p_3)
 f^{\mathrm{eq}}_{p_1}\bar{f}^{\mathrm{eq}}_{p_2}\bar{f}^{\mathrm{eq}}_{p_3}
 \nonumber\\
 &\times\big[\delta^3(\bs{p}-\bs{q})+\delta^3(\bs{p_1}-\bs{q})
 -\delta^3(\bs{p_2}-\bs{q})-\delta^3(\bs{p_3}-\bs{q})\big],
 \\[5pt]
 \label{eq:F}
 F_{ip}
 &\equiv F[\tilde{f}^{(i)}]_p
 \equiv -(f^{\mathrm{eq}}_p\bar{f}^{\mathrm{eq}}_p)^{-1}\left(\bs{v}\cdot\bs{\nabla}+\bs{F}\cdot\bs{\nabla}_p\right) \tilde{f}^{(i)}_p.
\end{align}
For arbitrary vectors $\psi_p$ and $\chi_p$, the linearized collision operator has the following three significant properties:
\begin{align}
 \left<\psi,L\chi\right> =\left<L\psi,\chi\right>, \ \ \ 
 \left<\psi,L\psi\right> \le 0, \ \ \  
 L\varphi^\alpha =0,
\end{align}
with the definition of the inner product given by
\begin{align}
 \left<\psi,\chi\right>\equiv 
 \int_p f^{\mathrm{eq}}_p\bar{f}^{\mathrm{eq}}_p\psi_p\chi_p.
\end{align}
The linearized collision operator has five zero modes,
\begin{align}
 \label{eq:zeromodes}
 \varphi_{0p}^0=1,\ \ \ 
 \varphi_{0p}^i=\delta p^i,\ \ \ 
 \varphi_{0p}^4=\frac{|\bs{\delta p}|^2}{2m}-h_\mathrm{TF},
\end{align}
with $h_\mathrm{TF}$ being the enthalpy density, the explicit form of which will be given shortly.
The zero modes satisfy the orthogonality relation
\begin{align}
\langle \varphi_0^\alpha, \varphi_0^\beta \rangle = c^\alpha\delta^{\alpha\beta}.
\end{align}
with the following normalization factors
\begin{align}
 c^0&=T\frac{\partial n}{\partial\,\mu_\mathrm{TF}},\ \ \ 
 c^i=mnT, \ \ \ 
 c^4=\frac{3nT}{2}\left(h_\mathrm{TF}-\frac{3n}{2}\left(\frac{\partial n}{\partial \mu_\mathrm{TF}}\right)^{-1}\right).
\end{align}
The particle number density $n$ and the enthalpy density $h_\mathrm{TF}$ are defined by
\begin{align}
 \label{eq:n}
 n(t,\bs{x}) 
 &\equiv \int_p f^{\mathrm{eq}}_p(t,\bs{x}),
 \\
 h_\mathrm{TF}(t,\bs{x})  
 &\equiv h(t,\bs{x})  - V(\bs{x}) 
 \equiv e(t,\bs{x})  + \frac{P(t,\bs{x}) }{n(t,\bs{x}) } -V(\bs{x}) .
 \\
\end{align}
The energy density $e$ and the pressure $P$ are given by
\begin{align}
 \label{eq:e}
 e(t,\bs{x}) 
 &\equiv \frac{1}{n}\int_p f^{\mathrm{eq}}_p(t,\bs{x}) \left(\frac{|\bs{\delta p}|^2}{2m} + V(\bs{x})\right),
 \\
 \label{eq:P}
 P(t,\bs{x}) 
 &\equiv \frac{1}{3}\int_p f^{\mathrm{eq}}_p(t,\bs{x}) \bs{\delta v}\cdot \bs{\delta p},
\end{align}
respectively.
We define P$_0$-space and Q$_0$-space as the space spanned by the zero modes given 
in Eq.~\eqref{eq:zeromodes} and its complemental space, respectively. 
The associated projection operators are defined as
\begin{align}
 \label{eq:P0}
 [P_0\psi]_p &\equiv \sum_{\alpha=0}^4\frac{\varphi^\alpha_{0p}}{c^\alpha}\langle\varphi_0^\alpha,\psi\rangle,
 \\
 \label{eq:Q0}
 Q_0 &\equiv 1-P_0,
\end{align} 
for an arbitrary vector $\psi_p$.
By solving the first order equation we obtain the first-order perturbative solution:
\begin{align}
 \label{eq:1st-psol1}
 \tilde{f}^{(1)}(t;t_0)=&f^{\mathrm{eq}}\bar{f}^{\mathrm{eq}}
 \Big[\mathrm{e}^{(t-t_0)L}(\Psi(t_0)+L^{-1}Q_0F_0)
 +(t-t_0)P_0F_0-L^{-1}Q_0F_0\Big].
\end{align}
Though we have formally obtained the first-order solution, the initial condition has not yet been determined.
The determination of $\Psi$ is done in the following way \cite{Tsumura:2015fxa}: 
We expand Eq.~\eqref{eq:1st-psol1} up to the first-order with respect to $|t-t_0|$,
\begin{align}
 \label{eq:1st-psol2}
 \tilde{f}^{(1)}(t;t_0)=&f^{\mathrm{eq}}\bar{f}^{\mathrm{eq}}
 \Big[\Psi(t_0)
 +(t-t_0)(L\Psi(t_0)+Q_0F_0+P_0F_0)\Big].
\end{align}
Such an expansion makes sense because we need only tangents of perturbative solution when applying the RG equation.
Now, we require that the tangent space of the perturbative solution becomes smallest, 
which is realized by choosing $L\Psi(t_0)$ so that $L\Psi(t_0)$ and $Q_0F_0$ belong 
to a common space, and we define P$_1$-space and Q$_1$-space 
by the space spanned by Eq.~\eqref{eq:1st-psol2} and its complemental space, respectively.
As is worked out in App.~\ref{sec:app1}, $L^{-1}Q_0F_0$ is calculated to be
\begin{align}
 \label{eq:LQF}
 [L^{-1}Q_0 F_0]_p= - \frac{\sigma^{ij}}{T}[L^{-1}\hat{\pi}^{ij}]_p - \frac{\nabla^i T}{T^2}[L^{-1}\hat{J}^i]_p,
\end{align}
where we have introduced excited modes $(\hat{\pi}^{ij}_p, \hat{J}^i_p)$ defined by
\begin{align}
 \hat{\pi}^{ij}_p &\equiv \delta v^{\langle i}\delta p^{j\rangle},
 \ \ \ 
 \hat{J}^i_p \equiv \left(\frac{|\bs{\delta p}|^2}{2m}-h_\mathrm{TF}\right)\delta v^i.
\end{align}
Here for an arbitrary tensor $A$, $A^{\langle ij\rangle}\equiv \Delta^{ijkl}A^{ij}$ 
with the definition of a symmetric traceless tensor 
$\Delta^{ijkl}\equiv \frac{1}{2}\delta^{ik}\delta^{jl}+
\frac{1}{2}\delta^{il}\delta^{jk}-\frac{1}{3}\delta^{ij}\delta^{kl}$. 
Therefore, it is physically natural to choose $\Psi(t_0)$ so that it belongs to the space spanned by $([L^{-1}\hat{\pi}^{ij}]_p, L^{-1}\hat{J}^i]_p)$, 
and $\Psi(t_0)$ may be parametrized as
\begin{align}
 \Psi_p(t_0) = -5\frac{[L^{-1}\hat{\pi}^{ij}]_p}{\langle\hat{\pi}^{kl},L^{-1}\hat{\pi}^{kl}\rangle}\pi^{ij}(t_0) 
 -3 \frac{[L^{-1}\hat{J}^{i}]_p}{\langle\hat{J}^{l},L^{-1}\hat{J}^{k}\rangle}J^{k}(t_0),
\end{align}
with eight would-be integral constants $\pi^{ij}(t_0)$ and $J^i(t_0)$, which will be identified as the stress tensor and heat flow, respectively.
From Eq.~\eqref{eq:1st-psol2}, P$_1$-space is found to be spanned by doublet modes $(\hat{\pi}^{ij}_p, \hat{J}^i_p)$ and $([L^{-1}\hat{\pi}^{ij}]_p, [L^{-1}\hat{J}^i]_p)$. The doublet modes are the very excited modes necessary for describing the mesoscopic dynamics.

The second-order equation reads
\begin{align}
 \frac{\partial}{\partial t}\tilde{f}^{(2)}(t;t_0)
 &=f^{\mathrm{eq}}\bar{f}^{\mathrm{eq}}L(f^{\mathrm{eq}}\bar{f}^{\mathrm{eq}})^{-1}\tilde{f}^{(2)}(t;t_0)
 +f^{\mathrm{eq}}\bar{f}^{\mathrm{eq}}K(t-t_0),
 \label{1st-peq2}
\end{align}
with the definitions,
\begin{align}
 K(t-t_0) &\equiv F[\tilde{f}^{(1)}(t)]
 +\frac{1}{2}B
 \left[(f^{\mathrm{eq}}\bar{f}^{\mathrm{eq}})^{-1}\tilde{f}^{(1)}(t),(f^{\mathrm{eq}}\bar{f}^{\mathrm{eq}})^{-1}\tilde{f}^{(1)}(t)\right],
 \\
 B[\chi,\psi]_{pq_1q_2} &\equiv -(f_v^{\mathrm{eq}}\bar{f}_p^{\mathrm{eq}})^{-1}
 \left.\frac{\delta^2}{\delta f_{q_1}\delta f_{q_2}}C[f]_v\right|_{f=f^{\mathrm{eq}}}
 f_{q_1}^{\mathrm{eq}}\bar{f}_{q_1}^{\mathrm{eq}}\chi_{q_1} f_{q_2}^{\mathrm{eq}}\bar{f}_{q_2}^{\mathrm{eq}}\psi_{q_2}.
\end{align}
Then, the second-order perturbative solution is given by
\begin{align}
 \tilde{f}^{(2)}(t;t_0)
 &=f^{\mathrm{eq}}\bar{f}^{\mathrm{eq}}
 \Bigg[\left(1-\mathrm{e}^{(t-t_0)\frac{\partial}{\partial s}}\right)
 \left(-\frac{\partial}{\partial s}\right)^{-1}P_0
 -\mathrm{e}^{(t-t_0)\frac{\partial}{\partial s}}
 \mathcal{G}(s)Q_0\Bigg]K(s)\Big|_{s=0},
\end{align}
with an initial condition
\begin{align}
 f^{(2)}(t_0)&=-f^{\mathrm{eq}}\bar{f}^{\mathrm{eq}}
 \mathcal{G}(s)Q_0 K(s)\Big|_{s=0},
\end{align}
which is chosen so as to exclude the fast dynamics in the Q$_1$-space.
Here
\begin{align}
 \mathcal{G}(s) \equiv \left(L-\frac{\partial}{\partial s}\right)^{-1}.
\end{align}

Thus, 
the perturbative solution up to the second order with respect to $\epsilon$ takes the following form:
\begin{align}
 \label{eq:1st-psol3}
 \tilde{f}(t;t_0)
 &=f^{\mathrm{eq}}
 +\epsilon f^{\mathrm{eq}}\bar{f}^{\mathrm{eq}}
 \Big[\mathrm{e}^{(t-t_0)L}(\Psi(t_0)+L^{-1}Q_0F_0)
 +(t-t_0)P_0F_0-L^{-1}Q_0F_0\Big]
 \nonumber\\
 &+\epsilon^2f^{\mathrm{eq}}\bar{f}^{\mathrm{eq}}
 \Bigg[\left(1-\mathrm{e}^{(t-t_0)\frac{\partial}{\partial s}}\right)
 \left(-\frac{\partial}{\partial s}\right)^{-1}P_0
 -\mathrm{e}^{(t-t_0)\frac{\partial}{\partial s}}
 \mathcal{G}(s)Q_0\Bigg]K(s)\Big|_{s=0},
\end{align}
with the initial condition:
\begin{align}
 \label{eq:initial2}
 f(t_0)=&f^{\mathrm{eq}}+\epsilon f^{\mathrm{eq}}\bar{f}^{\mathrm{eq}}\Psi
 -\epsilon^2f^{\mathrm{eq}}\bar{f}^{\mathrm{eq}}
 \mathcal{G}(s)Q_0 K(s)\Big|_{s=0}.
\end{align}

We have completed the perturbative calculation, but Eq.~\eqref{eq:1st-psol3} 
is valid only near the arbitrary initial time $t_0$ and apparently breaks down for large $|t-t_0|$ due to the secular terms.
To improve the perturbative solution into the global solution, we apply the RG equation given by
\begin{align}
 \label{eq:RGeq}
 \left.\frac{\mathrm{d}}{\mathrm{d}t_0}\right|_{t_0=t}\tilde{f}(t;t_0)=0,
\end{align}
which make the thirteen would-be integration constants 
$T(t_0)$, $u^i(t_0)$, $\mu(t_0)$, $\pi^{ij}(t_0)$, and $J^i(t_0)$ 
into the time-dependent dynamical variables  $T(t)$, $u^i(t)$, $\mu(t)$, $\pi^{ij}(t)$, and $J^i(t)$. 
In other words, the RG equation describes the dynamics of the hydrodynamic variables and the dissipative currents.
By substituting the solution of Eq.~\eqref{eq:RGeq} into $\tilde{f}(t;t_0=t)$,
we obtain the approximate solution to the original equation \eqref{eq:BE1} that has a validity in the global domain 
of time up to O$(\epsilon^2) $\cite{Kunihiro:1995zt}.
Geometrically speaking, the RG equation~\eqref{eq:RGeq} together with this substitution 
makes an envelope of the family of curves 
$\{\tilde{f}(t; t_0)\}_{t_0}$ parametrized by the arbitrary initial time $t_0$ \cite{Kunihiro:1995zt}. 
It can be shown that 
the envelope function thus constructed satisfies the Boltzmann equation in the global domain up to $\epsilon^2$.
In practice, we shall see that projections of the RG equation Eq.~\eqref{eq:RGeq} onto P${}_0$-space 
and P${}_1$-space indeed lead to the equation of continuity 
and the equation of relaxation, respectively \cite{Tsumura:2013uma,Tsumura:2015fxa}.

\subsection{
Hydrodynamics
}

The projection of the RG equation \eqref{eq:RGeq} onto the P${_0}$-space is done by taking the inner product of the zero modes \eqref{eq:zeromodes} and the RG equation \eqref{eq:RGeq} as follows,
\begin{align}
 \label{eq:averaging0}
 &\int_p\varphi^\alpha_{0p}
 \left[\frac{\partial}{\partial t} + \epsilon\left(\bs{v}\cdot \bs{\nabla} + \bs{F}\cdot \bs{\nabla}_p\right)\right][f^{\mathrm{eq}}+\epsilon f^{\mathrm{eq}}\bar{f}^{\mathrm{eq}}\Psi]_p
 = 0 + O(\epsilon^3).
\end{align}
While the inner product of the excited modes \eqref{eq:zeromodes} and the RG equation leads to 
\begin{align}
 \label{eq:averaging1}
 &\int_p [L^{-1}(\hat{\pi}^{ij},\hat{J}^i)]_p
 \left[\frac{\partial}{\partial t} + \epsilon\left(\bs{v}\cdot \bs{\nabla} + \bs{F}\cdot \bs{\nabla}_p\right)\right][f^{\mathrm{eq}}+\epsilon f^{\mathrm{eq}}\bar{f}^{\mathrm{eq}}\Psi]_p
 \nonumber\\
 &= \epsilon\langle L^{-1}(\hat{\pi}^{ij},\hat{J}^i), L\Psi \rangle
 +\epsilon^2\frac{1}{2}\langle L^{-1}(\hat{\pi}^{ij},\hat{J}^i), B[\Psi,\Psi] \rangle
 + O(\epsilon^3).
\end{align}
A straightforward calculation reduces Eq.~\eqref{eq:averaging0}  into the familiar equations
\begin{align}
 \label{eq:cont-eq-n3}
 \frac{\mathrm{D}n}{\mathrm{D}t} &= -n\boldsymbol{\nabla}\cdot\boldsymbol{u},
 \\
 \label{eq:cont-eq-u3}
 mn\frac{\mathrm{D}u^i}{\mathrm{D}t} &= -\nabla^i P + nF^i + \nabla^j \pi^{ij},
 \\
 \label{eq:cont-eq-s3}
 Tn\frac{\mathrm{D}s}{\mathrm{D}t} &= \boldsymbol{\nabla}\cdot\boldsymbol{J} + \sigma^{jk}\pi^{jk},
\end{align}
where we have introduced the Lagrange derivative defined by $\mathrm{D}/\mathrm{D}t=\partial/\partial t+\bs{u}\cdot\bs{\nabla}$.
As worked out in Appendix.~\ref{sec:app2}, Eq.~\eqref{eq:averaging1} is reduced into the following 
relaxation equation\com{s}
\begin{align}
 \label{eq:relax1}
  \pi^{ij}
  &= 2\eta\sigma^{ij}
  - \tau_\pi \frac{\mathrm{D}}{\mathrm{D}t}\pi^{ij}
  - \ell_{\pi J}\nabla^{\langle i} J^{j\rangle} 
  \nonumber\\
  &+ \kappa_{\pi\pi}^{(1)}\pi^{ij}\boldsymbol{\nabla}\cdot\boldsymbol{u}
 + \kappa_{\pi\pi}^{(2)}\pi^{k\langle i}\sigma^{j\rangle k}
 - 2\tau_\pi \pi^{k\langle i}\omega^{j\rangle k}
 \nonumber\\
 &+ \kappa_{\pi J}^{(1)}J^{\langle i}\nabla^{j\rangle}n
 + \kappa_{\pi J}^{(2)}J^{\langle i}\nabla^{j\rangle}P
 + \kappa_{\pi J}^{(3)}J^{\langle i}F^{j\rangle}
  \nonumber\\
  &+ b_{\pi\pi\pi} \pi^{k\langle i} \pi^{j\rangle k}
  + b_{\pi JJ} J^{\langle i} J^{j\rangle},
\\[10pt]
  \label{eq:relax2}
 J^i
  &=\lambda \nabla^i T
  -\tau_J \frac{\mathrm{D}}{\mathrm{D}t}J^i
  -\ell_{J\pi}\nabla^j \pi^{ij}
  \nonumber\\
  &+ \kappa_{J\pi}^{(1)}\pi^{ij}\nabla^j n
 + \kappa_{J\pi}^{(2)}\pi^{ij}\nabla^j P
 + \kappa_{J\pi}^{(3)}\pi^{ij}F^j
 \nonumber\\
 &+ \kappa_{JJ}^{(1)}J^i\boldsymbol{\nabla}\cdot\boldsymbol{u}
 + \kappa_{JJ}^{(2)}J^j\sigma^{ij}
 + \tau_J J^j\omega^{ij}
  \nonumber\\
  &+ b_{JJ\pi}J^j \pi^{ij}.
\end{align}
The microscopic expressions of the shear viscosity and heat conductivity, which are the first-order transport coefficients, are given by
\begin{align}
 \label{eq:TC1}
 \eta &\equiv -\frac{1}{10T}\langle\hat{\pi}^{ij},L^{-1}\hat{\pi}^{ij}\rangle,
 \\
 \label{eq:TC2}
 \lambda &\equiv -\frac{1}{3T^2}\langle\hat{J}^i,L^{-1}\hat{J}^i\rangle,
\end{align}
while the viscous relaxation times are given by
\begin{align}
 \label{eq:RT1}
 \tau_{\pi} &\equiv \frac{1}{10T\eta}\langle\hat{\pi}^{ij},L^{-2}\hat{\pi}^{ij}\rangle,
 \\
 \label{eq:RT2}
 \tau_{J} &\equiv \frac{1}{3T^2\lambda}\langle\hat{J}^{i},L^{-2}\hat{J}^{i}\rangle.
\end{align}
Explicit expressions of the other coefficients are summarized in App.~\ref{sec:app2}.
Defining ``time-evolved" vectors $\{\hat{\pi}^{ij}_p(s),\hat{J}^{ij}_p(s)\} = \{[\mathrm{e}^{sL}\hat{\pi}^{ij}]_p,[\mathrm{e}^{sL}\hat{J}^{ij}]_p\}$, we may convert Eqs.~\eqref{eq:TC1}-\eqref{eq:RT2} into the following forms
\begin{align}
 \label{eq:TC}
 &\eta = \frac{1}{10T}\int_0^\infty\mathrm{d}s\langle\hat{\pi}^{ij}(0),\hat{\pi}^{ij}(s)\rangle,
 \ \ \ 
 \lambda = \frac{1}{3T^2}\int_0^\infty\mathrm{d}s\langle\hat{J}^i(0),\hat{J}^i(s)\rangle,
 \\
 \label{eq:RT}
 &\tau_{\pi} = \frac{\int_0^\infty\mathrm{d}s\,s\langle\hat{\pi}^{ij}(0),\hat{\pi}^{ij}(s)\rangle}{\int_0^\infty\mathrm{d}s\langle\hat{\pi}^{ij}(0),\hat{\pi}^{ij}(s)\rangle},
 \ \ \ 
 \tau_{J} = \frac{\int_0^\infty\mathrm{d}s\,s\langle\hat{J}^i(0),\hat{J}^i(s)\rangle}{\int_0^\infty\mathrm{d}s\langle\hat{J}^i(0),\hat{J}^i(s)\rangle},
\end{align}
which may give the clearer physical interpretation. 
Some remarks are in order here:
(i)~Eq.~\eqref{eq:TC} 
is consistent to the Green-Kubo formula \cite{Jeon:1994if,Jeon:1995zm,Hidaka:2010gh}.
They actually take the same form as those of the Chapman-Enskog method.
(ii)~The forms of Eq.~\eqref{eq:RT} give the time constant 
of the correlation function of the microscopic dissipative currents, 
which are physically natural forms as viscous relaxation times.
(iii)~As is already expressed in the relaxation equations,
the transport coefficient of $\pi^{k\langle i}\omega^{j\rangle k}$ 
in Eq.~\eqref{eq:relax1} and that of $J^i\omega^{ij}$ in Eq.~\eqref{eq:relax2}
identically coincide with $-2\tau_\pi$ and $\tau_J$, respectively;
the analytical derivation of these relations are given in Appendix.~\ref{sec:app2}.
(iv)~Terms including quadratic vorticity term $\omega^{k\langle i}\omega^{j\rangle k}$ 
do not appear in our relaxation equations.

The last two points are consistent with the result shown in Ref.~\cite{Schafer:2014}. 
We should emphasize that these results are obtained in an exact manner without
recourse to any approximation.

%%%%%%%%%%%%%%%%%%%%%%%%%%%%%%%%%%%%%%%%%%%%%%%%%%%%%%%%%%%%%%%%%%%%%

%%%%%%%%%%%%%%%%%%%%%%%%%%%%%%%%%%%%%%%%%%%%%%%%%%%%%%%%%%%%%%%%%%%%%
%\setcounter{equation}{0}
\section{
Computational method of the transport coefficients and relaxation times}
\label{sec:sec4}

In this section, 
bearing the application to the ultracold Fermi gases realized in the cold-atom experiments in mind, 
we compute  the transport coefficients and the relaxation times
to calculate the shear viscosity, heat conductivity, 
and the viscous relaxation times of the stress tensor and heat flow of cold Fermi gasses
assuming that the s-wave scattering is dominant in the collision integral \eqref{eq:collision}:
To consider the s-wave scattering in the Fermi gases, we assume the gases are 
composed of spin-1/2 particles and they interact in the singlet state.
Then, the scattering amplitude in Eq.~\eqref{eq:transition} is given by
\begin{align}
 \label{eq:cross_section}
 \mathcal{M} = \frac{4\pi}{a_s^{-1}-iq},
\end{align}
where $a_s$ is the s-wave scattering length and $\bs{q}=(\bs{p}-\bs{p}_1)/2$ is the incoming relative momentum.
Given the scattering amplitude Eq.~\eqref{eq:cross_section}, 
the reduction of the expressions \eqref{eq:TC1}-\eqref{eq:RT2} to 
forms workable for numerical computations is still involved.
%The numerical calculation of the first-order transport coefficients 
%and the viscous relaxation times given by Eqs.~\eqref{eq:TC1}-\eqref{eq:RT2} 
%consists of the following three steps 
By adapting the method developed for the relativistic case \cite{de1980relatlvlatlc},
we make the reduction in the following three steps:
\begin{itemize}
 \item Simplify the linearized collision operator \eqref{eq:LCO} 
 by performing integrations with Eq.~\eqref{eq:cross_section} being inserted.
 \item Discretize the momentum with a finite number $N_p$ of momenta, and
 solve the finite-dimensional linear equation $\big[ L X \big]_p = (\hat{\pi}^{ij},\hat{J}^i)_p$ 
to obtain the $N_p$-dimensional vector $X_p \equiv \big[ \hat{L}^{-1}(\hat{\pi}^{ij},\hat{J}^i) \big]_p$ numerically.
 \item Evaluate Eqs.~\eqref{eq:TC1}-\eqref{eq:RT2} with $X_p = \big[ \hat{L}^{-1}(\hat{\pi}^{ij},\hat{J}^i) \big]_p$ 
thus obtained inserted numerically.
\item Increase the number $N_p$, and repeat the above procedure until the convergence is achieved.
\end{itemize}

%As we discussed in Sec.~\ref{sec:sec4}, 
First we introduce the following dimensionless quantities for later covenience:
\begin{align}
 &\bs{\tau} \equiv \frac{\bs{\delta p}}{p_F} = \frac{\bs{\delta v}}{v_F},
 \\
 &\mu' \equiv \frac{\mu_{TF}}{\varepsilon_F},
 \ \ \
 T' \equiv \frac{T}{\varepsilon_F},
 \\
 &n' \equiv \frac{n}{p_F^3} = 1/(3\pi^2),
\end{align}
with the Fermi velocity and the Fermi energy
\begin{align}
 p_F\equiv (3\pi^2 n)^{1/3},\ \ \ 
 v_F\equiv \frac{p_F}{m},\ \ \ 
 \varepsilon_F \equiv \frac{p_F^2}{2m}.
\end{align}
$\bs{\tau}$ is a dimensionless momentum and
the primed variables are also dimensionless. We shall suppress the prime in this appendix.

First, we reduce the linearized collision operator.
A dimensionless linearized collision operator is given by
\begin{align}
 \label{eq:L}
 L'[\phi]_\tau
 &\equiv \frac{1}{4\varepsilon_F}L[\phi]_p
 =-\frac{1}{2\bar{f}^\mathrm{eq}_\varepsilon}\int_{\tau_1}\int_{\tau_2}\int_{\tau_3}
 \mathcal{W}'(\tau,\tau_1|\tau_2,\tau_3)
  f^\mathrm{eq}_{\tau_1}\bar{f}^\mathrm{eq}_{\tau_2}\bar{f}^\mathrm{eq}_{\tau_3}
 (\phi_p+\phi_{p_1}-\phi_{p_2}-\phi_{p_3}).
\end{align}
The equilibrium distribution function is given by
\begin{align}
 f^\mathrm{eq}_\varepsilon = \frac{1}{\mathrm{e}^{(\varepsilon-\mu)/T}-a},
\end{align}
with the dimensionless energy density $\varepsilon=|\bs{\tau}|^2\equiv \tau^2$.
The dimensionless transition matrix $\mathcal{W}'$ reads
\begin{align}
 \mathcal{W}'(\tau,\tau_1|\tau_2,\tau_3)&=|\mathcal{M}'|^2(2\pi)^4\delta(\tau^2+\tau_1^2-\tau_2^2-\tau_3^2)
 \delta^3(\bs{\tau}+\bs{\tau}_1-\bs{\tau}_2-\bs{\tau}_3)
\end{align}
with the dimensionless scattering amplitudes defined by $\mathcal{M}'\equiv mp_F\mathcal{M}$.
For the s-wave scattering, the dimensionless scattering amplitude is given by
\begin{align}
 |\mathcal{M}'|^2 \equiv |mp_F\mathcal{M}|^2
 =\frac{16\pi^2}{(p_F a_s)^{-2}+q^2},
\end{align}
where we have introduced $2\bs{q}\equiv (\bs{p}-\bs{p}_1)/p_F=\bs{\tau}-\bs{\tau}_1$.
In this paper, we suppose that the scattering amplitude depends on only $q=|\boldsymbol{q}|$, i.e.,
$\mathcal{M}=\mathcal{M}(q)$.

Now, we convert the linearized collision operator \eqref{eq:L} into a convenient form for the numerical calculations.

\begin{align}
 \label{eq:L1}
 L'[\phi]_\tau
 &=-k(\bs{\tau})\phi_p-\int_{\tau_1} f^\mathrm{eq}_{\varepsilon_1}[K_1(\bs{\tau},\bs{\tau}_1)-K_2(\bs{\tau},\bs{\tau}_1)]\phi_{p_1}
\end{align}
with the following definitions
\begin{align}
 \label{eq:int1}
 k(\bs{\tau})
 &=\int_{\tau_1} f^\mathrm{eq}_{\varepsilon_1}K_1(\bs{\tau},\bs{\tau}_1),
 \\
 \label{eq:int2}
 K_1(\bs{\tau},\bs{\tau}_1)
 &=\frac{1}{2\bar{f}^\mathrm{eq}_\varepsilon}\int_{\tau_2}\int_{\tau_3}
 |\mathcal{M}'(|\bs{\tau}-\bs{\tau}_1|/2)|^2
 \bar{f}^\mathrm{eq}_{\varepsilon_2}\bar{f}^\mathrm{eq}_{\varepsilon_3}
 \nonumber\\
 &\times(2\pi)^4\delta(\tau^2+\tau_1^2-\tau_2^2-\tau_3^2)
 \delta^{3}(\bs{\tau}+\bs{\tau}_1-\bs{\tau}_2-\bs{\tau}_3),
 \\
 \label{eq:int3}
 K_2(\bs{\tau},\bs{\tau}_1)
 &=\frac{\mathrm{e}^{(\varepsilon_1-\mu)/T}}{\bar{f}^\mathrm{eq}_\varepsilon}\int_{\tau_2}\int_{\tau_3}
 |\mathcal{M}'(|\bs{\tau}-\bs{\tau}_2|/2)|^2
 f^\mathrm{eq}_{\varepsilon_2}\bar{f}^\mathrm{eq}_{\varepsilon_3}
 \nonumber\\
 &\times(2\pi)^4\delta(\tau^2+\tau_2^2-\tau_1^2-\tau_3^2)
 \delta^{3}(\bs{\tau}+\bs{\tau}_2-\bs{\tau}_1-\bs{\tau}_3).
\end{align}
We partially perform the integrations \eqref{eq:int1}-\eqref{eq:int3}.
To this end, we define the total momentum and relative momentum as follows,
\begin{align}
 \label{eq:velocity1}
    \bs{P}\equiv \bs{\tau}+\bs{\tau}_1,
    \ \ \ 
    \bs{q}\equiv (\bs{\tau}-\bs{\tau}_1)/2,
    \ \ \ 
    \bs{P}'\equiv \bs{\tau}_2+\bs{\tau}_3,
    \ \ \ 
    \bs{q}'\equiv (\bs{\tau}_2-\bs{\tau}_3)/2.
\end{align}
Then, Eq.~\eqref{eq:int2} may be converted as follows:
\begin{align}
 K_1(\bs{p},\bs{p}_1)
 &=\frac{1}{2\bar{f}^{\mathrm{eq}}_\varepsilon}\int_{P'}\int_{q'} |\mathcal{M}'(q)|^2
 \bar{f}^\mathrm{eq}_{|\bs{P}'+2\bs{q}'|^2/4}
 \bar{f}^\mathrm{eq}_{|\bs{P}'-2\bs{q}'|^2/4}
 \nonumber\\
 &\times (2\pi)^4\delta\left(\frac{P^2}{2}+2q^2-\frac{P^{'2}}{2}-2q^{'2}\right)
 \delta^{3}(\bs{P}-\bs{P}')
 \nonumber\\
 &=\frac{q}{16\pi\bar{f}^\mathrm{eq}_\varepsilon}|\mathcal{M}'(q)|^2
 \int\mathrm{d}\cos\theta
 \bar{f}^{\mathrm{eq}}_{P^2/4+q^2+Pq\cos\theta}
 \bar{f}^{\mathrm{eq}}_{P^2/4+q^2-Pq\cos\theta}
 \nonumber\\
 &=\frac{q}{16\pi\bar{f}^{\mathrm{eq}}_\epsilon}|mp_F\mathcal{M}(q)|^2
 \frac{T}{\left(1-a^2\mathrm{e}^{-(P^2+4q^2-4\mu)/2T}\right)Pq}
 \nonumber\\
 &\times\ln\left(\frac{\left(1-a\mathrm{e}^{-(P^2+4q^2+4Pq-4\mu)/4T}\right)
 \left(\mathrm{e}^{(P^2+4q^2+4Pq-4\mu)/4T}-a\right)}
 {\left(1-a\mathrm{e}^{-(P^2+4q^2-4Pq-4\mu)/4T}\right)
 \left(\mathrm{e}^{(P^2+4q^2-4Pq-4\mu)/4T}-a\right)}\right),
\end{align}
where
\begin{align}
 \cos\theta&=\frac{\bs{P}\cdot\bs{q}'}{Pq'},
 \\
 P&=\sqrt{\tau^2 +2\tau\tau_1\cos\chi+\tau_1^2},
 \\
 q&=\frac{\sqrt{\tau^2 -2\tau\tau_1\cos\chi+\tau_1^2}}{2},
 \\
 \cos\chi&=\frac{\bs{\tau}\cdot\bs{\tau}_1}{\tau\tau_1}.
\end{align}
So we may write as
\begin{align}
 K_1(\bs{\tau},\bs{\tau}_1)=K_1(\tau,\tau_1,\chi).
\end{align}
Therefore, Eq.~\eqref{eq:int1} reads
\begin{align}
 k(\bs{\tau})
 &=\frac{1}{(2\pi)^2}\int\mathrm{d}\tau_1\tau_1^2 f^\mathrm{eq}_{\varepsilon_1}
 \int\mathrm{d}\cos\chi K_1(\tau,\tau_1,\chi)
 \equiv k(\tau).
\end{align}

\begin{figure}[t]
 \begin{center}
 \includegraphics[width=4cm]{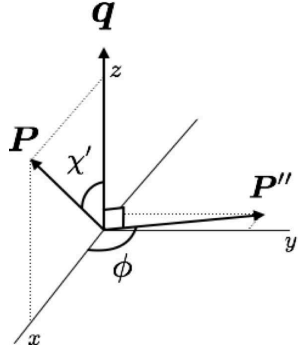}
 \caption{Schematic figure}
 \label{fig1}
 \end{center}
\end{figure}

Next, we evaluate Eq.~\eqref{eq:int3}. 
\begin{align}
 K_2(\bs{\tau},\bs{\tau}_1)
 &=\frac{\mathrm{e}^{(\varepsilon_1-\mu)/T}}{\bar{f}^\mathrm{eq}_\varepsilon}\int_{P'}\int_{q'}
 |\mathcal{M}'(|\bs{P}-\bs{P}'+2\bs{q}-2\bs{q}'|/4)|^2
 f^\mathrm{eq}_{|\bs{P}'+2\bs{q}'|^2/4}
 \bar{f}^\mathrm{eq}_{|\bs{P}'-2\bs{q}'|^2/4}
 \nonumber\\
 &\times(2\pi)^4\delta(2\bs{P}\cdot\bs{q}+2\bs{P}'\cdot\bs{q}')
 \delta^{(3)}(2\bs{q}+2\bs{q}')
 \nonumber\\
 &=\frac{\mathrm{e}^{(\varepsilon_1-\mu)/T}}{64\pi^2q \bar{f}^\mathrm{eq}_\varepsilon}\int\mathrm{d}P''P''\mathrm{d}\phi
 \big|\mathcal{M}'\big(\sqrt{q^2+P''^2/16}\big)\big|^2
 \nonumber\\
 &\times f^\mathrm{eq}_{(|\bs{P}-2\bs{q}|^2+P''^2+2\bs{P}''\cdot\bs{P})/4}
 \bar{f}^\mathrm{eq}_{(|\bs{P}+2\bs{q}|^2+P''^2+2\bs{P}''\cdot\bs{P})/4}
 \nonumber\\
 &\equiv K_2(\tau,\tau_1,\chi),
\end{align} 
where  we have changed the integration variables from $\bs{P}'$ to $\bs{P}''=\bs{P}'-\bs{P}$.
The inner product $\bs{P}''\cdot\bs{P}$ is evaluated as follows:
We can take the Cartesian coordinates such that the vectors $\bs{P}''$ and $\bs{P}$ are parametrized as
\begin{align}
 \bs{P}= P\,{}^t(\sin\chi', 0, \cos\chi'),\ \ \ 
 \bs{P}''= P''\,{}^t(\cos\phi, \sin\phi, 0),
\end{align}
without loss of generality (see Fig.~\ref{fig1}). 
Then, the inner product $\bs{P}''\cdot\bs{P}$ is evaluated as
\begin{align}
 \bs{P}''\cdot\bs{P} = PP''\sin\chi'\cos\phi,
\end{align}
with
\begin{align}
 \cos\chi' \equiv \frac{\bs{P}\cdot\bs{q}}{Pq}.
\end{align}

%%%%%%%%%%%%%%%%%%%%%%%%%%%%%%%%%%%%%%%%%%%%%%%%%

%%%%%%%%%%%%%%%%%%%%%%%%%%%%%%%%%%%%%%%%%%%%%%%%%

Secondly, by using the expression of $L$ obtained above, we evaluate $[L^{-1}\hat{\pi}^{ij}]_p$ and $[L^{-1}\hat{J}^i]_p$ in the following way.
Microscopic expressions of the stress tensor and the heat flow are given by
\begin{align}
 \hat{\pi}^{ij}_p &= -p_Fv_F\frac{\sqrt{k(\tau)}}{\tau^2}c^*(\tau)\tau^{\langle i}\tau^{j\rangle},
 \ \ \ 
 \hat{J}^i_p = -\varepsilon_F v_F\frac{\sqrt{k(\tau)}}{\tau}b^*(\tau)\tau^i,
\end{align}
with the definitions
\begin{align}
 c^*(\tau) &\equiv -\frac{\tau^2}{\sqrt{k(\tau)}},
 \ \ \ 
 b^*(\tau) \equiv -\frac{\tau}{\sqrt{k(\tau)}}(\tau^2-h_\mathrm{TF}),
\end{align}
where $h_\mathrm{TF}$ is the (dimensionless) enthalpy given by
\begin{align}
 h_\mathrm{TF}=\frac{5}{2}\int\mathrm{d}\tau\,\tau^4 f^\mathrm{eq}_\varepsilon.
\end{align}
Without loss of generality, $L^{-1}\hat{\pi}^{ij}$ and $L^{-1}\hat{J}^i$ may be expressed in terms of scalars $C^*(\tau)$ and $B^*(\tau)$ as
\begin{align}
 \label{eq:C_B}
 [L'^{-1}\hat{\pi}^{ij}]_p &= \frac{p_Fv_F}{\tau^2\sqrt{k(\tau)}}C^*(\tau)\tau^{\langle i}\tau^{j\rangle},
 \ \ \ 
 [L'^{-1}\hat{J}^i]_p = \frac{\varepsilon_F p_F}{\tau\sqrt{k(\tau)}}B^*(\tau)\tau^i.
\end{align}
Then, $\hat{\pi}^{ij}_p=L'[L'^{-1}\hat{\pi}^{ij}]_p$ and $\hat{\pi}^{ij}_p=L'[L'^{-1}\hat{\pi}^{ij}]_p$ can be written as
\begin{align}
 c^*(\tau)
 =C^*(\tau)
 &+\frac{1}{(2\pi)^2}\int\mathrm{d}\tau_1\tau_1^2 f^\mathrm{eq}_{\varepsilon_1}\frac{1}{\sqrt{k(\tau)k(\tau_1)}}
 \nonumber\\
 &\times\int\mathrm{d}\cos\chi\left(\frac{3}{2}\cos^2\chi-\frac{1}{2}\right)
 [K_1(\tau,\tau_1,\chi)-K_2(\tau,\tau_1,\chi)]C^*(\tau_1),
 \\
 b^*(\tau)
 =B^*(\tau)
 &+\frac{1}{(2\pi)^2}\int\mathrm{d}\tau_1\tau_1^2 f^\mathrm{eq}_{\varepsilon_1}\frac{1}{\sqrt{k(\tau)k(\tau_1)}}
 \nonumber\\
 &\times\int\mathrm{d}\cos\chi\cos\chi
 [K_1(\tau,\tau_1,\chi)-K_2(\tau,\tau_1,\chi)]B^*(\tau_1).
\end{align}
By solving these equations numerically, we obtain $C^*(\tau)$ and $B^*(\tau)$, which gives $L^{-1}\hat{\pi}^{ij}$ and $L^{-1}\hat{J}^i$ by using Eq.~\eqref{eq:C_B}.

Finally, the transport coefficients may be obtained by evaluating the expressions given by
\begin{align}
 \label{eq:TC1a}
 \eta
 &=-\frac{1}{10(\varepsilon_F T)}\langle\hat{\pi}^{ij},L^{-1}\hat{\pi}^{ij}\rangle
 =n\frac{1}{10T}\int\mathrm{d}\tau\,\tau^2 f^\mathrm{eq}_\varepsilon \bar{f}^\mathrm{eq}_\varepsilon c^*(\tau)C^*(\tau),
 \\
 \label{eq:TC2a}
 \lambda
 &=-\frac{1}{3(\varepsilon_F T)^2}\langle\hat{J}^i,L^{-1}\hat{J}^i\rangle
 =\frac{n}{m}\frac{1}{4T^2}\int\mathrm{d}\tau\,\tau^2 f^\mathrm{eq}_\varepsilon \bar{f}^\mathrm{eq}_\varepsilon b^*(\tau)B^*(\tau),
\end{align}
for the first-order transport coefficients, and
\begin{align}
 \label{eq:RT1a}
 \tau_\pi
 &=\frac{1}{10(\varepsilon_F T)\eta}\langle L^{-1}\hat{\pi}^{ij},L^{-1}\hat{\pi}^{ij}\rangle
 =\frac{1}{\varepsilon_F}\frac{1}{40T(\eta/n)}\int\mathrm{d}\tau\,\tau^2f^\mathrm{eq}_\varepsilon 
 \bar{f}^\mathrm{eq}_\varepsilon \frac{1}{k(\tau)} C^*(\tau)C^*(\tau),
 \\
 \label{eq:RT2a}
 \tau_J
 &=\frac{1}{3(\varepsilon_F T)^2\lambda}\langle L^{-1}\hat{J}^i,L^{-1}\hat{J}^i\rangle
 =\frac{1}{\varepsilon_F}\frac{1}{16T^2(m\lambda/n)}\int\mathrm{d}\tau\,\delta p^2 f^\mathrm{eq}_\varepsilon \bar{f}^\mathrm{eq}_\varepsilon \frac{1}{k(\tau)} B^*(\tau)B^*(\tau),
\end{align}
for the viscous relaxation times.

In practice, the momentum is discretized with a degrees of freedom $N_p$, and we vary
$N_p$ until the convergence of the results is achieved. 
The convergence of the numerical results are shown in Sec.~\ref{sec:sec5-3}.

%%%%%%%%%%%%%%%%%%%%%%%%%%%%%%%%%%%%%%%%%%%%%%%%%%%%%%%%%%%%%%%%%%%%%

%%%%%%%%%%%%%%%%%%%%%%%%%%%%%%%%%%%%%%%%%%%%%%%%%%%%%%%%%%%%%%%%%%%%%
%\setcounter{equation}{0}
\section{
Numerical results}
\label{sec:sec5}

In this section, we show the numerical results
of  the shear viscosity, heat conductivity, 
and the viscous relaxation times of the stress tensor and heat flow 
of cold Fermi gasses. Varying the scattering length $a_s$ and temperature,
we focus on the quantum statistical effects of the transport coefficeints and relaxation times.
We also exaine the relaxation-time approximation in a quantitative way using 
our results; some of the results are briefly reported in Ref.~\cite{kikuchi:2015letter}.

%%%%%%%%%%%%%%%%%%%%%%%%%%%%%%%%%%%%%%%%%%%%%%
\subsection{Quantum statistical effects on the transport coefficients}

\begin{figure}[t]
 \begin{minipage}{0.45\hsize}
 \begin{center}
 \includegraphics[width=6.5cm]{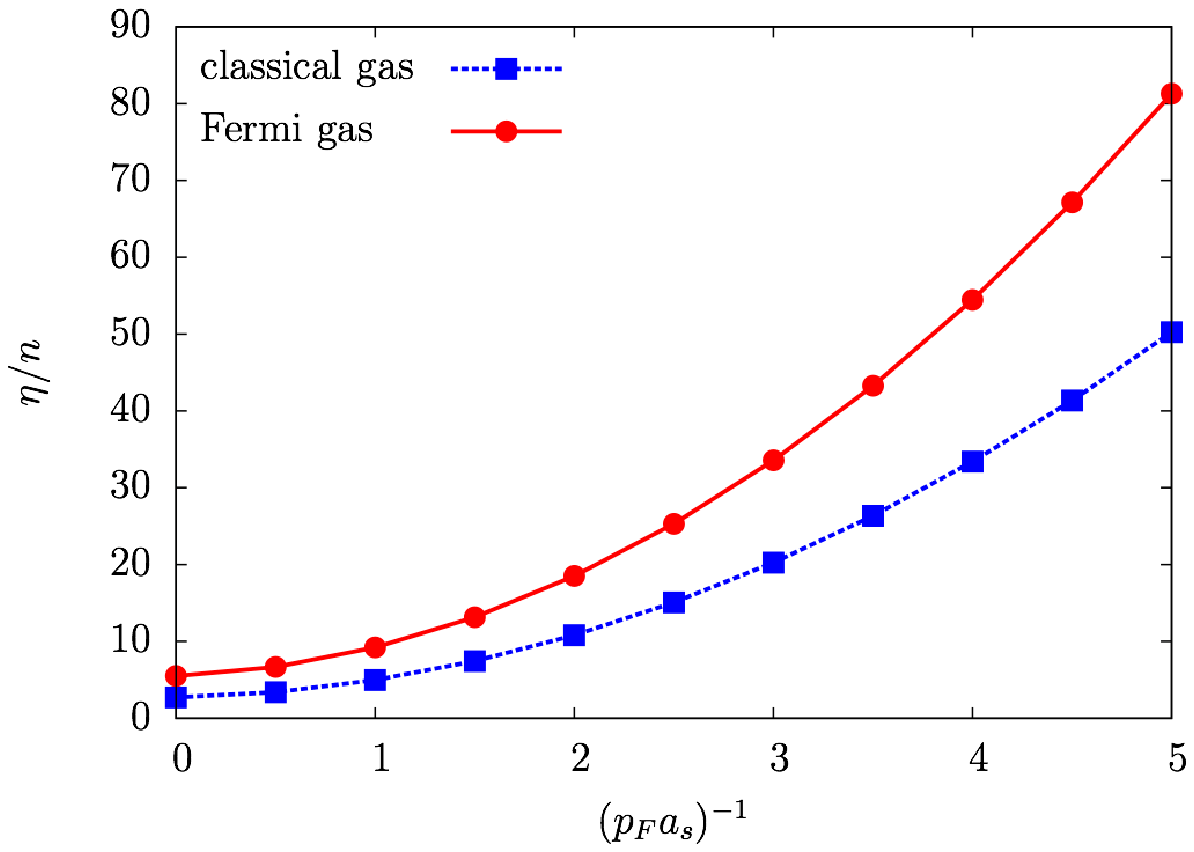}
 \end{center}
 \end{minipage}
 \begin{minipage}{0.4\hsize}
 \begin{center}
 \includegraphics[width=6.5cm]{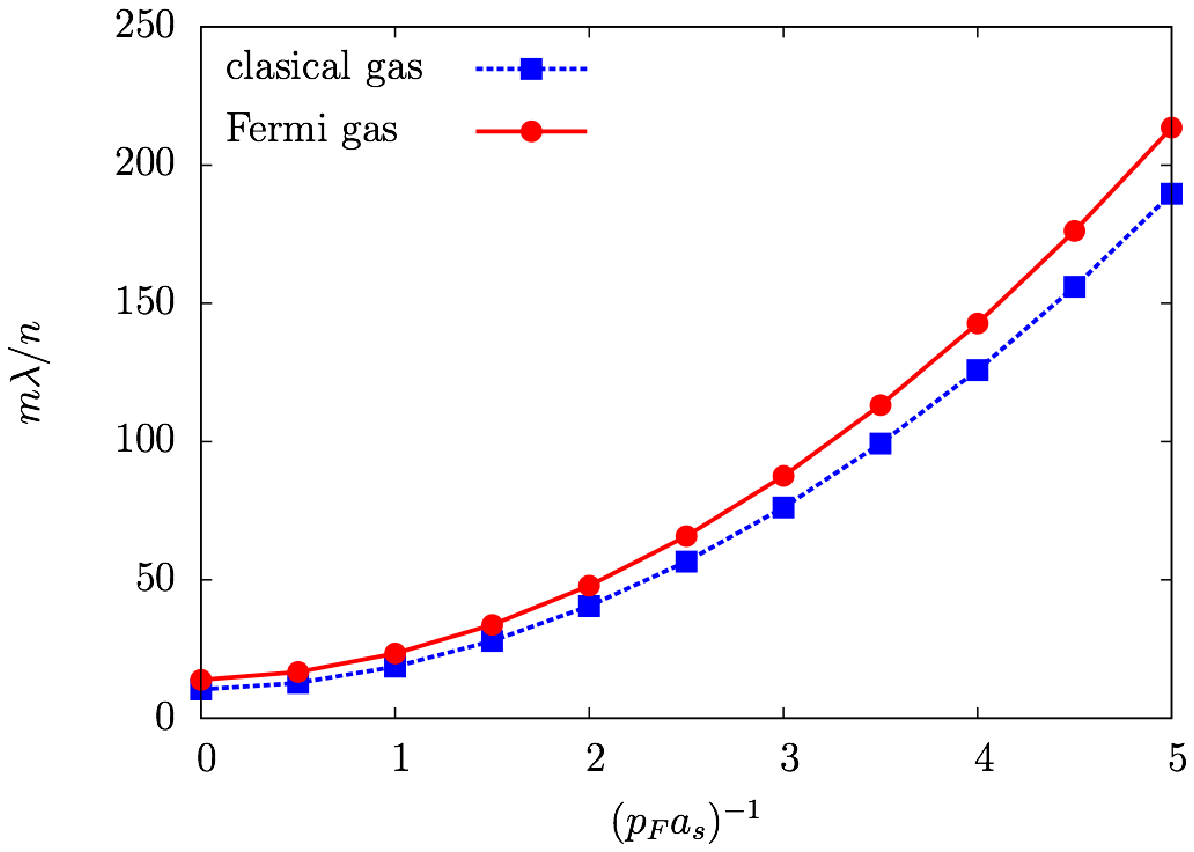}
 \end{center}
 \end{minipage}
 \caption{Scattering length dependence of the shear viscosity (left panel) and heat conductivity (right panel) at the Fermi temperature, $T=T_F$. The blue square and red circle indicate the classical Boltzmann gas and Fermi gas, respectively.}
 \label{fig:a_1st_CF}
\end{figure}

\begin{figure}[t]
 \centering
 \includegraphics[width=7cm]{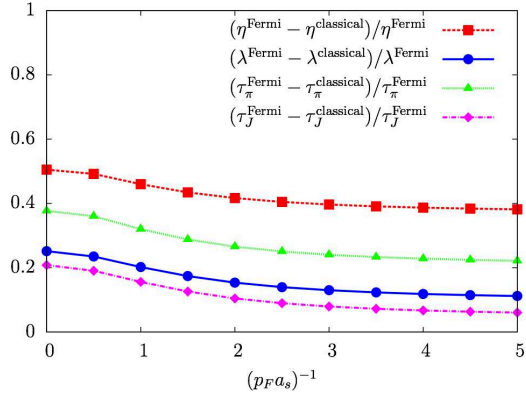}
 \caption{Scattering length dependence of the quantum statistical effects for the shear viscosity (red square), heat conductivity (blue circle), the viscous relaxation times of the stress tensor (green triangle), and that of the heat conductivity (purple rhombus) are respectively shown at the Fermi temperature.}
 \label{fig:ratio_Q}
\end{figure}

\begin{figure}[t]
 \begin{minipage}{0.45\hsize}
 \begin{center}
 \includegraphics[width=6.5cm]{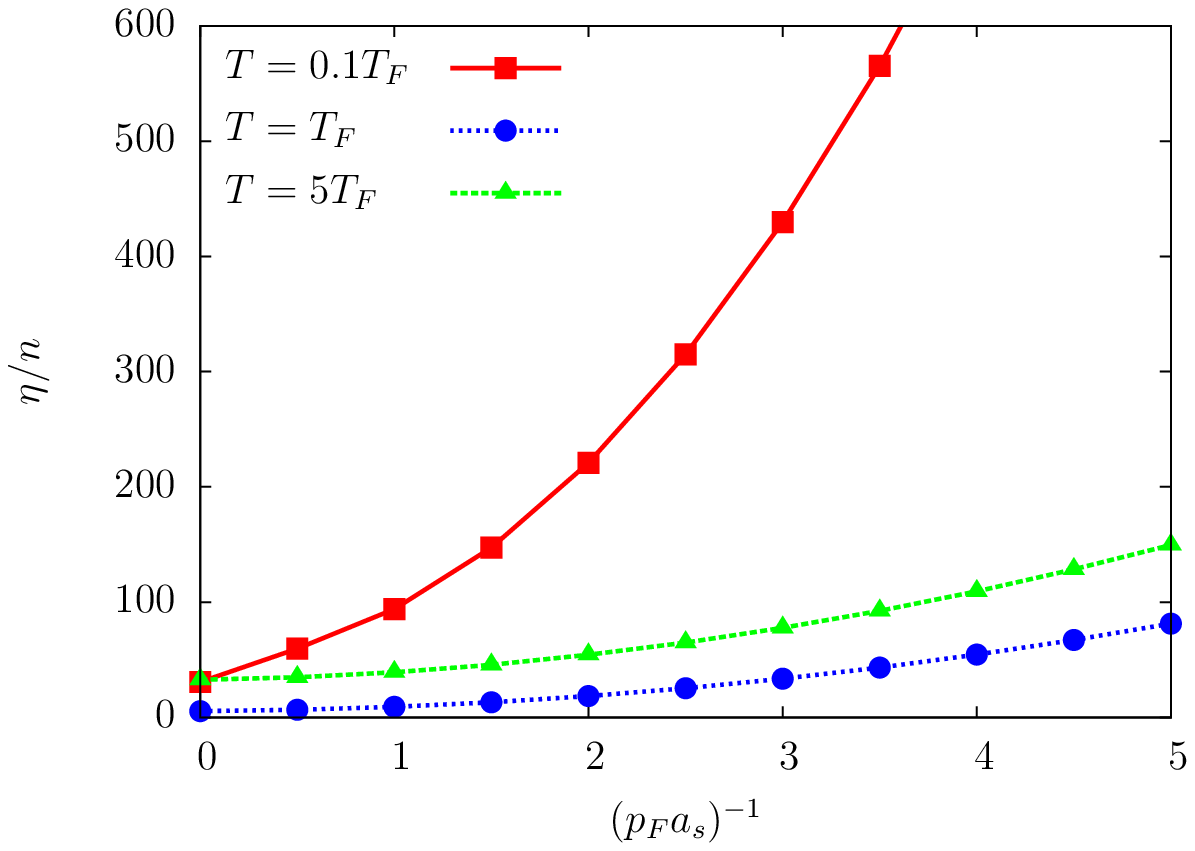}
 \end{center}
 \end{minipage}
 \begin{minipage}{0.45\hsize}
 \begin{center}
 \includegraphics[width=6.5cm]{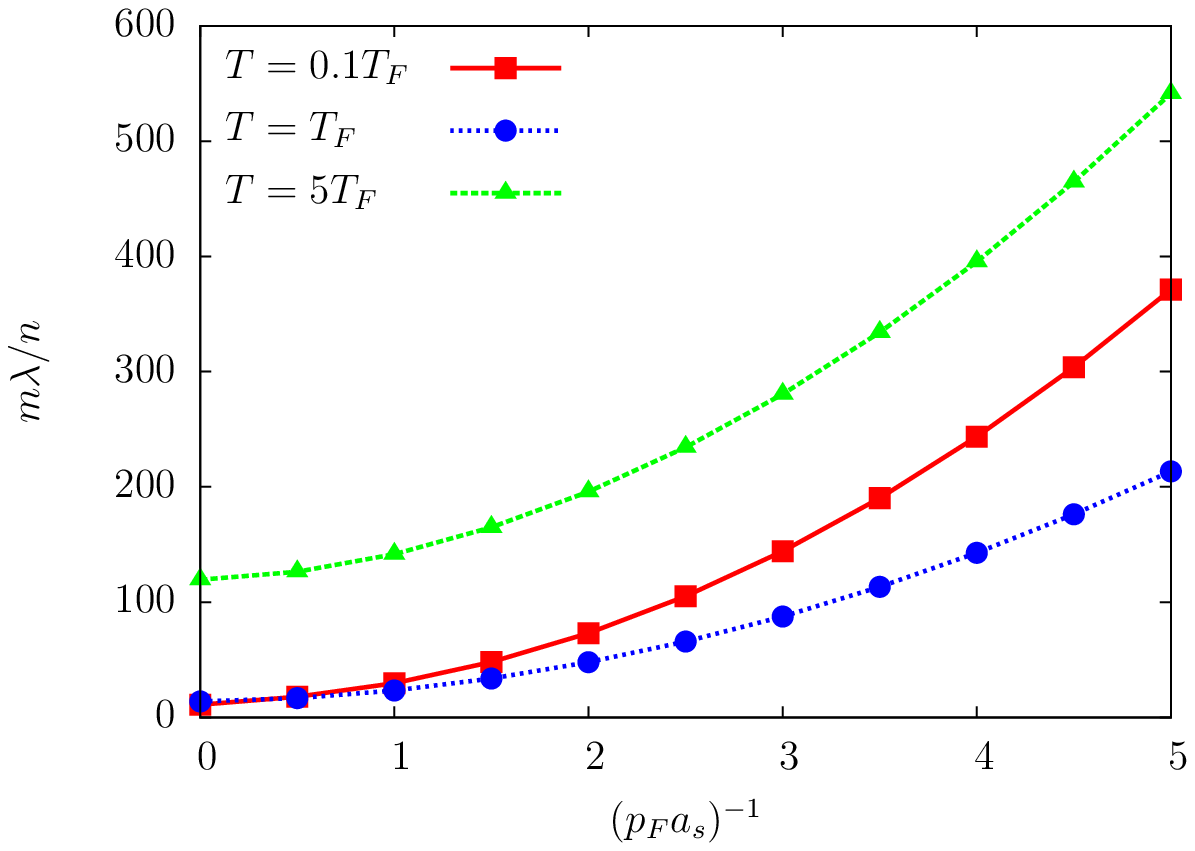}
 \end{center}
 \end{minipage}
 \caption{Scattering length dependence of the shear viscosity (left panel) and heat conductivity (right panel). The red square, blue circle, and green triangle correspond to the Fermi gas with the temperature $T/T_F=0.1,\ 1,\ 5$, respectively.}
 \label{fig:a_1st}
\end{figure}

First, we calculate the shear viscosity and heat conductivity with varying the scattering length. The unitary limit is 
taken by $a_s\to \infty$, i.e., $(p_F a_s)^{-1}=0$, where $p_F$ is the Fermi momentum. For the classical Boltzmann gas, we set $a=0$ in Eq.~\eqref{eq:collision} and the corresponding equilibrium distribution function reads $f^\mathrm{eq}_p=\mathrm{e}^{-(|\bs{\delta p}|^2/2m-\mu_\mathrm{TF})/T}$. For the Fermi gas, we take into account the quantum statistics by setting $a=-1$, and the equilibrium distribution function is $f^\mathrm{eq}_p=[\mathrm{e}^{(|\bs{\delta p}|^2/2m-\mu_\mathrm{TF})/T}+1]^{-1}$.
The resultant data show that, as the scattering length increases, the viscous effects decrease (Fig.~\ref{fig:a_1st_CF}), and  the quantum statistical effects increase (Fig.~\ref{fig:ratio_Q}). The scattering-length dependence of the quantum statistical effects indicates that the quantum nature becomes apparent at unitarity, where the atomic gases are strongly correlated.
We also see that the shear viscosity and heat conductivity decrease as the scattering length becomes smaller and they are smallest at the unitary limit for any temperature (Fig.~\ref{fig:a_1st}).

\begin{figure}[t]
 \begin{minipage}{0.45\hsize}
 \begin{center}
 \includegraphics[width=6.5cm]{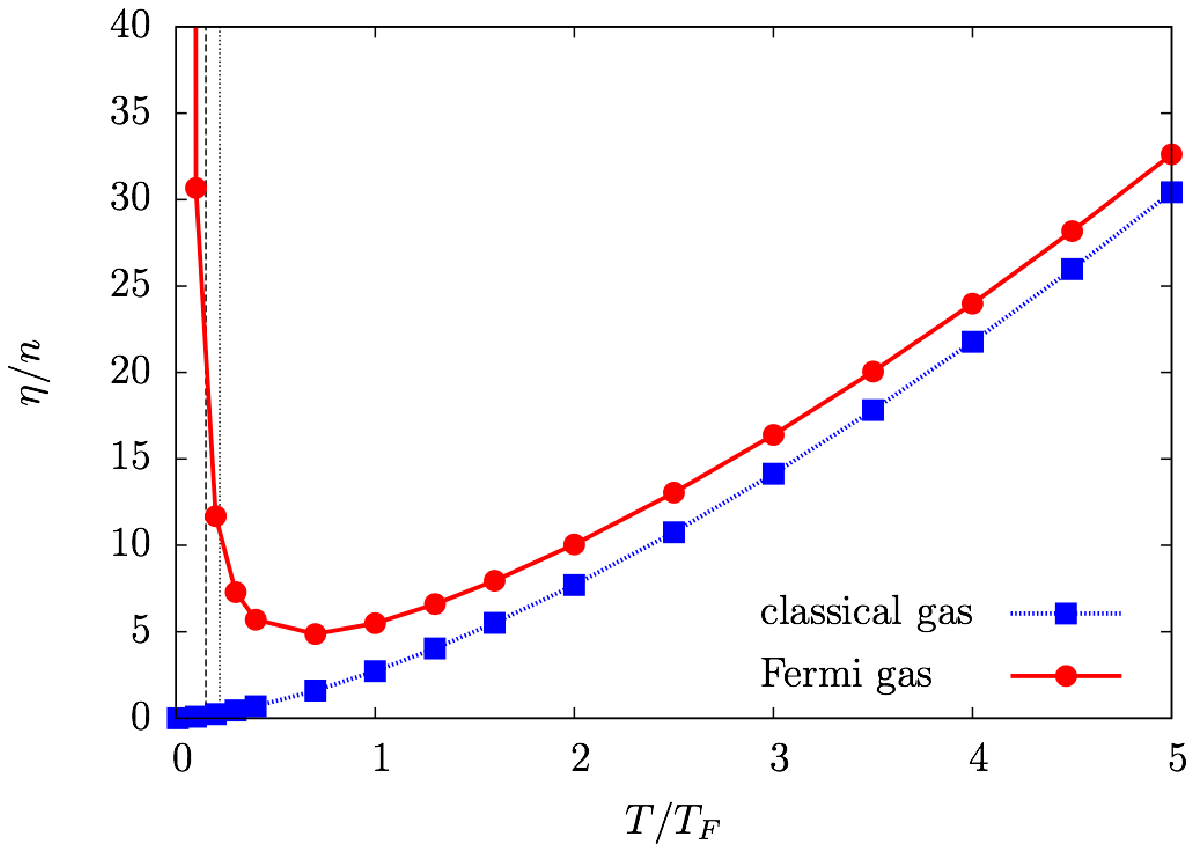}
 \end{center}
 \end{minipage}
 \begin{minipage}{0.45\hsize}
 \begin{center}
 \includegraphics[width=6.5cm]{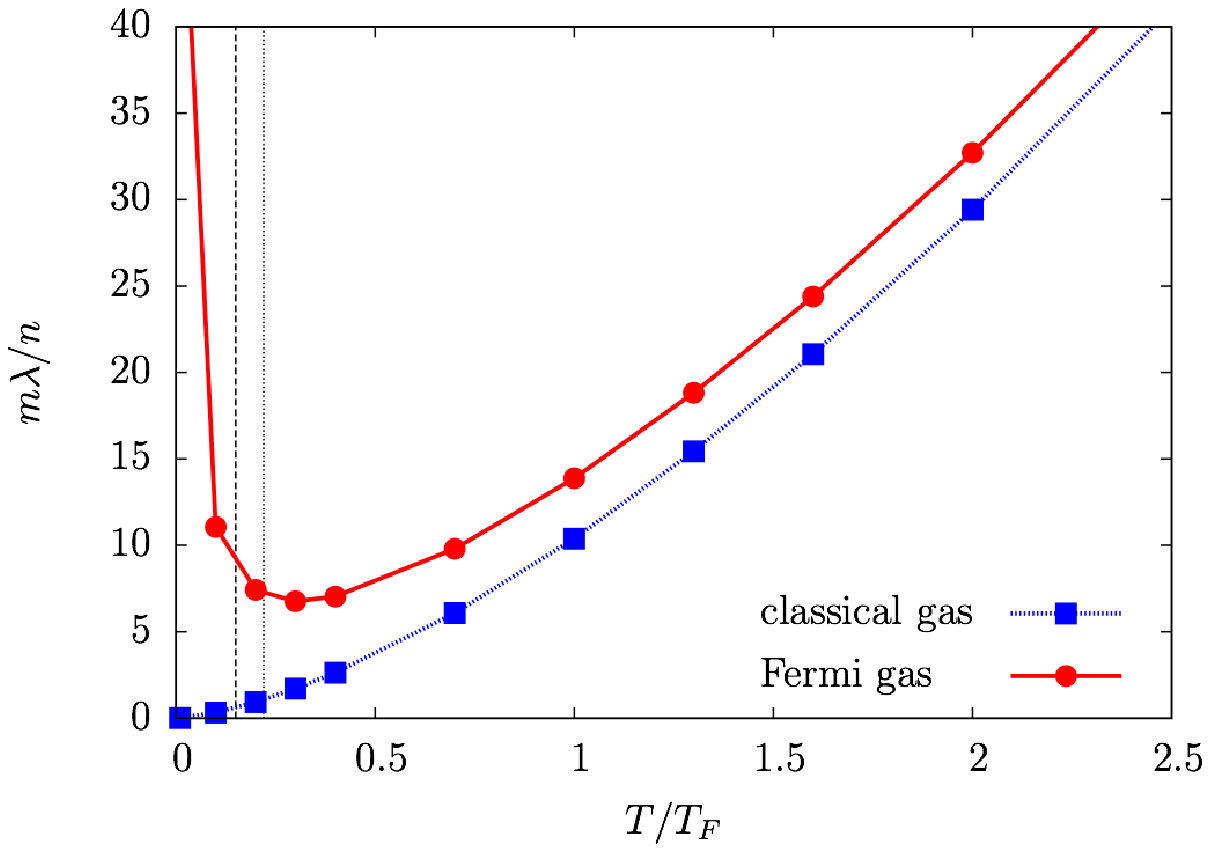}
 \end{center}
 \end{minipage}
 \caption{Temperature dependence of the shear viscosity (left panel) and heat conductivity (right panel) at the unitarity, $(p_Fa_s)^{-1}=0$. The blue square and red circle indicate the first-order transport coefficients of classical Boltzmann gas and the Fermi gas, respectively. The black dashed and dotted lines show the superfluid-transition temperature $T_c\simeq 0.15T_F$ and the pairing-formation temperature $T^*\simeq 0.22T_F$, respectively.
 }
 \label{fig:T_1st}
\end{figure}

Figure~\ref{fig:T_1st} shows the temperature dependence of $\eta/n$ and $m\lambda/n$. 
The quantum statistical effect increases the $\eta/n$ and $m\lambda/n$. The difference between the classical gas and the Fermi gas becomes larger as temperature decreases. In particular, at the temperature $T<T_F$, the differences become clear and those of the Fermi gases diverge.
Three comments are in order here.
(i)~The quantum statistical effect is not negligible even above $T^*$, 
below which pairing effects dominate and our computation becomes invalid.
(ii)~The increases of the shear viscosity and heat conductivity for the Fermi gas 
at small temperature are naturally understood in terms of the Pauli blocking effect;
a good Fermi sphere is formed at low temperature and thus the elastic scattering rate other than the forward one
is so greatly suppressed due to the Pauli blocking that the energy 
and momentum transport become quite efficient, which implies the increase of the shear viscosity and heat conductivity.
(iii)~Our result of the shear viscosity might implies that the kinetic approach is unreliable near the unitarity at low temperature even above the $T^*$, in contrast to the previous works which claim good agreement between experimental results and theoretical results based on the Boltzmann equation without quantum Fermi statistics as low as $0.3T_F$ at unitarity.

\begin{figure}[t]
 \begin{minipage}{0.45\hsize}
 \begin{center}
 \includegraphics[width=6.5cm]{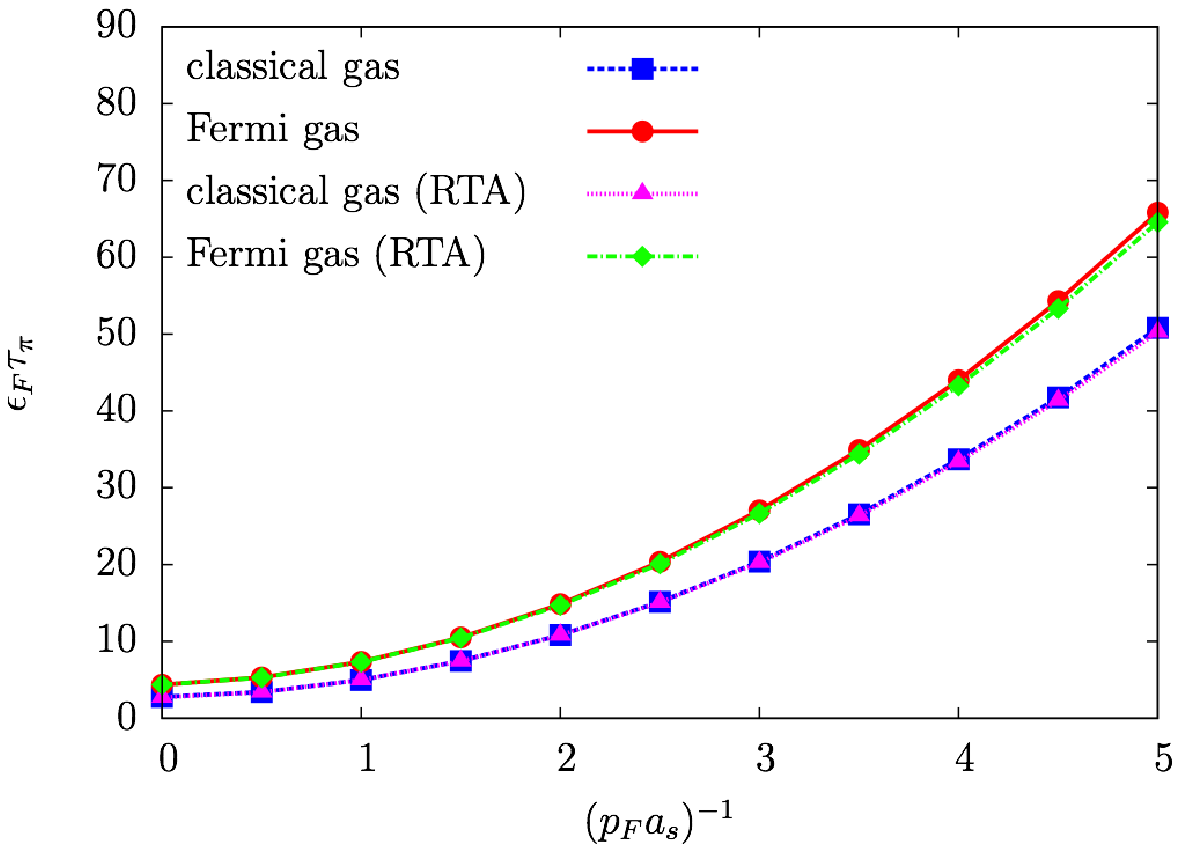}
 \end{center}
 \end{minipage}
 \begin{minipage}{0.45\hsize}
 \begin{center}
 \includegraphics[width=6.5cm]{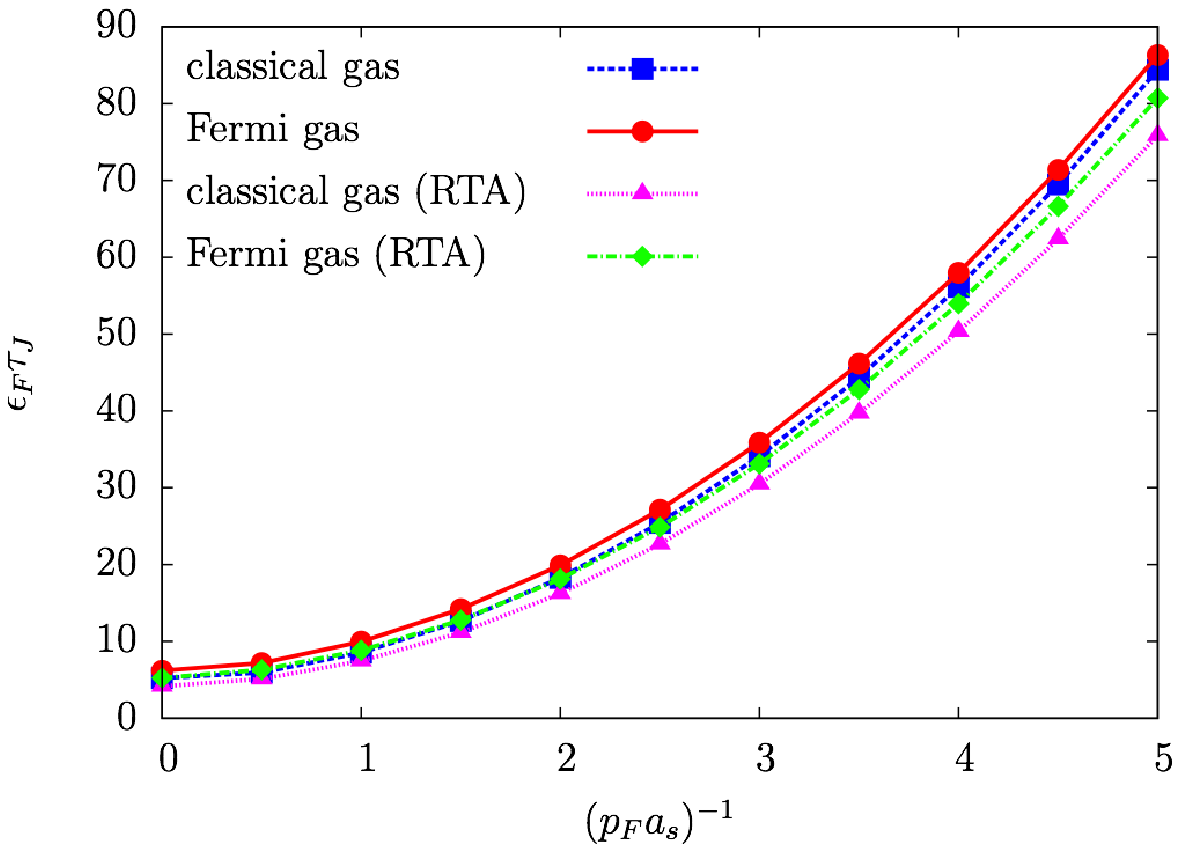}
 \end{center}
 \end{minipage}

 \caption{Scattering length dependence of the viscous relaxation times of the stress tensor (left panel) and heat flow (right panel) at the Fermi temperature, $T=T_F$. The blue square and red circle indicate the viscous relaxation times of classical Boltzmann gas and the Fermi gas which evaluated without the RTA, while the purple triangle and green rhombus indicate those evaluated with the RTA given by Eqs.~\eqref{eq:setting1} and \eqref{eq:setting2}. The black dashed and dotted lines show $T_c$ and $T^*$, respectively.}
 \label{fig:a_tau}
\end{figure}

\begin{figure}[t]
 \begin{minipage}{0.45\hsize}
 \begin{center}
 \includegraphics[width=6.5cm]{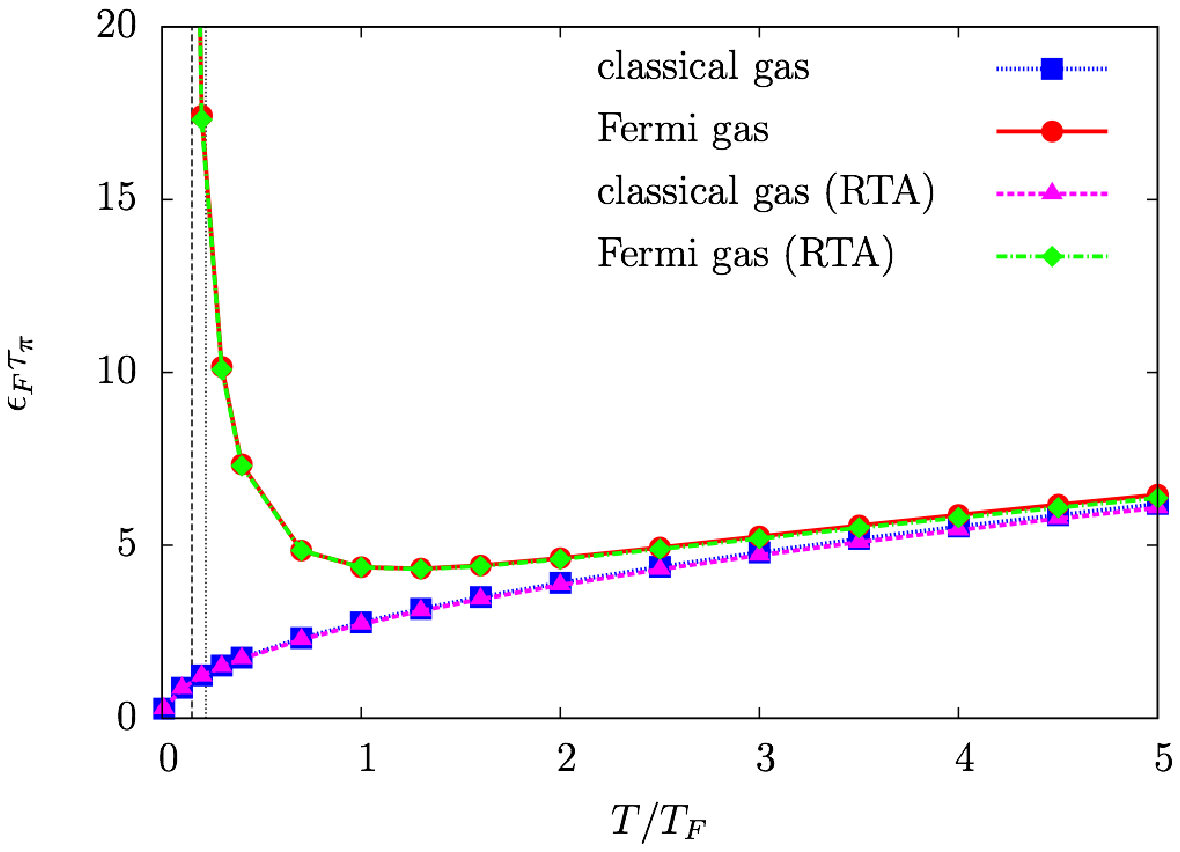}
 \end{center}
 \end{minipage}
 \begin{minipage}{0.45\hsize}
 \begin{center}
 \includegraphics[width=6.5cm]{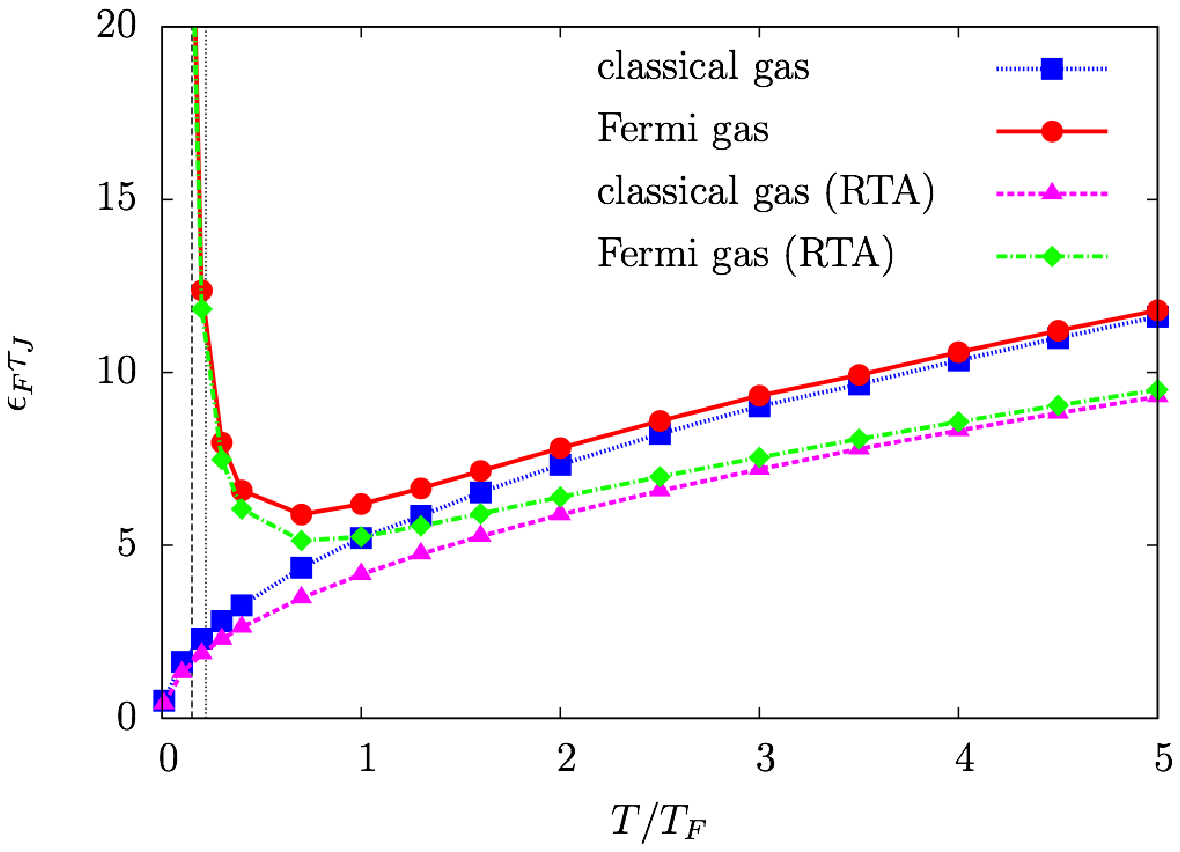}
 \end{center}
 \end{minipage}
 \caption{Temperature dependence of the viscous relaxation times of the stress tensor (left panel) and heat flow (right panel) at the unitarity, $(p_Fa_s)^{-1}=0$. The blue square and red circle indicate the viscous relaxation times of classical Boltzmann gas and the Fermi gas which evaluated without the RTA, while the purple triangle and green rhombus indicate those evaluated with the RTA given by Eqs.~\eqref{eq:setting1} and \eqref{eq:setting2}. The black dashed and dotted lines show $T_c$ and $T^*$, respectively.}
 \label{fig:T_tau}
\end{figure}

The viscous relaxation times exhibits similar behaviors as the first-order transport coefficients. The differences between the classical gas and the Fermi gas increase at low temperature and near unitarity (see Fig.~\ref{fig:a_tau} and \ref{fig:T_tau}). 

\begin{figure}[t]
 \centering
 \includegraphics[width=7cm]{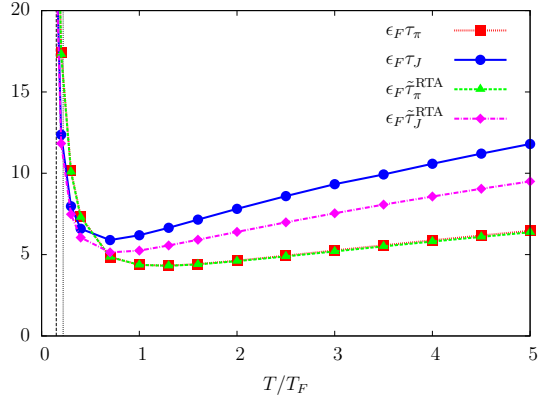}
 \caption{Temperature dependence of the viscous relaxation times of the stress tensor and heat flow of the unitary Fermi gas. The red square and blue circle respectively indicate the viscous relaxation times of the stress tensor and heat conductivity which evaluated without the RTA, while the green triangle and purple rhombus indicate those evaluated with the RTA given by Eqs.~\eqref{eq:setting1} and \eqref{eq:setting2}. The black dashed and dotted lines show $T_c$ and $T^*$, respectively.}
 \label{fig:T_tau2}
\end{figure}

\begin{figure}[t]
 \centering
 \includegraphics[width=7cm]{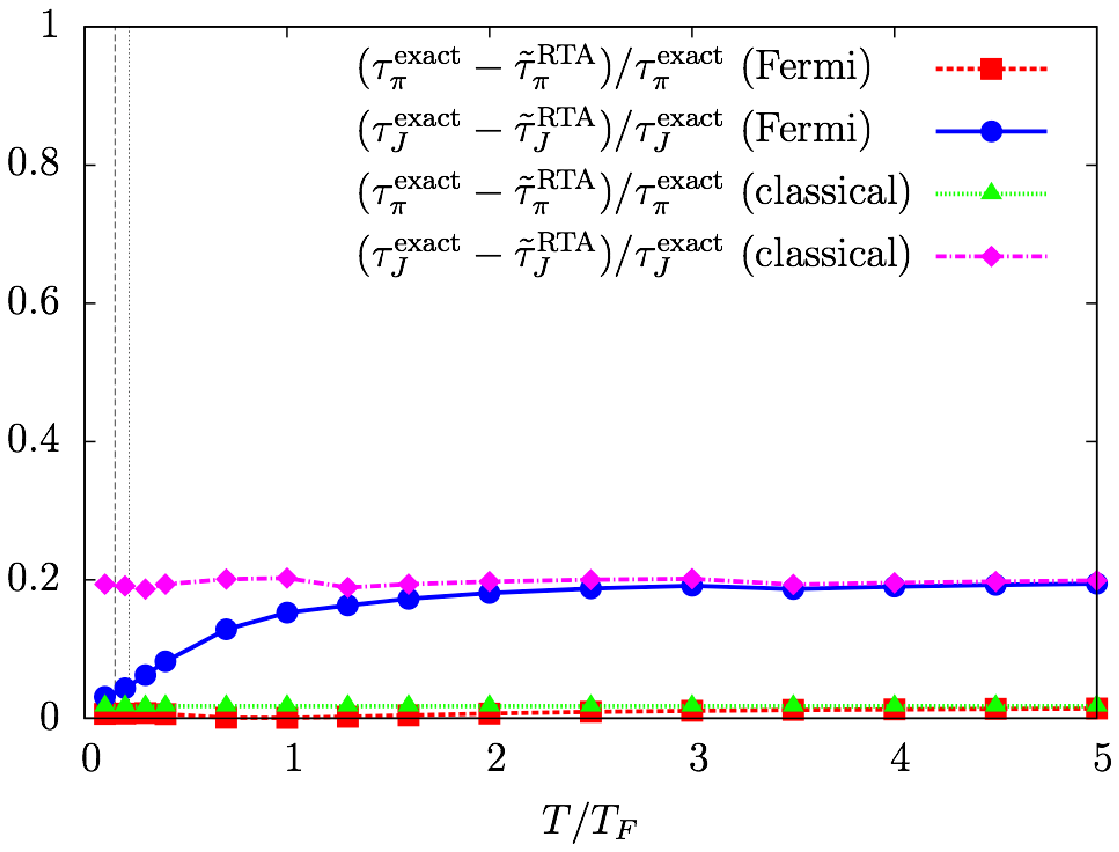}
 \caption{Temperature dependence of the error caused by the RTA are shown for the shear viscosity (red square), heat conductivity (blue circle), the viscous relaxation times of the stress tensor (green triangle), and that of the heat conductivity (purple rhombus) are respectively shown at unitarity. The black dashed and dotted lines show $T_c$ and $T^*$, respectively.}
 \label{fig:ratio_RTA}
\end{figure}

%%%%%%%%%%%%%%%%%%%%%%%%%%%%%%%%%%%%%%%%%%%%
\subsection{Reliability of the relaxation-time approximation}

In the RTA, the collision integral in Eq.~\eqref{eq:collision} is replaced by
\begin{align}
 C[f]_p = -\frac{f_p-f^\mathrm{eq}_p}{\tau},
\end{align}
where $\tau$ is a free parameter which determines the time scale for a non-equilibrium system to relax toward the equilibrium states. 
Under the approximation, the shear viscosity and heat conductivity are calculated to be \cite{Bruun:2007,Braby:2011,Chao:2012}
\begin{align}
 \label{eq:transport_rel}
 \eta^{\mathrm{RTA}} = \tau P,
 \ \ \ 
 \lambda^{\mathrm{RTA}} = \frac{\tau}{12mT}\left(7Q-\frac{75P^2}{n}\right),
\end{align}
where $Q$ is defined by $Q \equiv \int_p \delta v^2 \delta p^2 f^\mathrm{eq}_p$.
In addition, it should be noted that the viscous relaxation times are given by $\tau$,
\begin{align}
 \label{eq:tau_rel}
 \tau^{\mathrm{RTA}}_\pi = \tau^{\mathrm{RTA}}_J = \tau.
\end{align}
In the RTA, the relaxation-time scales of the system is characterized by only one parameter $\tau$, which must be determined phenomenologically or based on more elaborated microscopic analyses.
However, the viscous relaxation times of the stress tensor and heat conductivity given by Eqs.~\eqref{eq:RT1} and \eqref{eq:RT2} take the considerably different values (Fig.~\ref{fig:T_tau2}).
This results are clearly contradict to Eq.~\eqref{eq:tau_rel} and indicate that the RTA should be modified so as to incorporate the multiple relaxation-time scales.
To this end, we determine the viscous relaxation times independently with the help of the relations Eq.~\eqref{eq:transport_rel} derived from the RTA and exact value of the shear viscosity and heat conductivity as follows:
We evaluate the viscous relaxation time of the stress tensor by
\begin{align}
 \label{eq:setting1}
 \tau = \frac{\eta^{\mathrm{exact}}}{P} \equiv \tilde{\tau}_\pi^\mathrm{RTA},
\end{align}
While the viscous relaxation time of the heat conductivity is evaluated as
\begin{align}
 \label{eq:setting2}
 \tau = \frac{12mT\lambda^{\mathrm{exact}}}{(7Q-75P^2/n)} \equiv \tilde{\tau}_\pi^\mathrm{RTA}.
\end{align}
We note that
$\eta^{\mathrm{exact}}$, $\lambda^{\mathrm{exact}}$, $\tau^{\mathrm{exact}}_\pi$, and $\tau^{\mathrm{exact}}_J$
denote the transport coefficients and viscous relaxation times,
which are respectively calculated from Eqs.~\eqref{eq:TC1}-\eqref{eq:RT2}
in an exact manner, as in the analyses in the last subsection.
In Fig.~\ref{fig:a_tau} and \ref{fig:T_tau}, we compare the viscous relaxation times with and without the RTA
and the numerical results of Eqs.~\eqref{eq:setting1} and \eqref{eq:setting2} actually behave similarly compared with the exact ones qualitatively.
It is remarkable that
$\eta^{\mathrm{exact}}/P$
well reproduces the viscous relaxation time of the stress tensor
$\tau^{\mathrm{exact}}_\pi$
for both the classical gas and the Fermi gas regardless of temperature and scattering length. On the other hand, the RTA still has the quantitative error and always underestimates the viscous relaxation time of the heat conductivity compared with those evaluated exactly.
It is noteworthy that the error of $\tau_J$ caused by the RTA for the Fermi gas decreases at low temperature in contrast to that for classical gas, which is independent of temperature (Fig.~\ref{fig:ratio_RTA}).
This behavior may be understood as follows:
The RTA is a kind of the linear approximation of the collision integral with respect to the deviation of the distribution function from the equilibrium one.
The deviation is attributed to the excitations of quasi-particles due to the nonequilibrium process and decreases in the cold fermionic gases 
since the Pauli-blocking effect suppresses the excitation of the quasi-particles inside the Fermi sphere.
Therefore the linear approximation work well and the RTA reproduces the exact value of the viscous relaxation time of the heat conductivity.

It is noted that the expressions \eqref{eq:setting1} and \eqref{eq:setting2} 
derived with the RTA can be obtained by a closure approximation
\begin{align}
\label{eq:replace_RTA}
\langle\hat{\pi}^{ij},L^{-2}\hat{\pi}^{ij}\rangle \to \langle\hat{\pi}^{ij},L^{-1}\hat{\pi}^{ij}\rangle \langle\hat{\pi}^{ij},
L^{-1}\hat{\pi}^{ij}\rangle/\langle\hat{\pi}^{ij},\hat{\pi}^{ij}\rangle,
\end{align}
in the exact expressions given by Eqs.~\eqref{eq:RT1} and \eqref{eq:RT2}.
Such a closure approximation is reminiscent of the mean-field (or Hartree) approximation, and
it may imply that the RTA might be validated if the vector space $L\hat{\pi}^{ij}$ is saturated by $\hat{\pi}^{ij}$.
Equation~\eqref{eq:setting1} is verified by noticing $\langle\hat{\pi}^{ij},\hat{\pi}^{ij}\rangle = 10TP$ and $\langle\hat{\pi}^{ij},L^{-1}\hat{\pi}^{ij}\rangle = 10T\eta^\mathrm{exact}$ from Eq.~\eqref{eq:TC1}.
Then by applying the replacement~\eqref{eq:replace_RTA} to Eq.~\eqref{eq:RT1}, we obtain Eq.~\eqref{eq:setting1} as,
\begin{align}
 &\tau_{\pi} = \frac{1}{10T\eta^\mathrm{exact}}\langle\hat{\pi}^{ij},L^{-2}\hat{\pi}^{ij}\rangle
 \nonumber\\
 &\to
 \tilde{\tau}^\mathrm{RTA}_{\pi} = \frac{1}{10T\eta^\mathrm{exact}}\frac{\langle\hat{\pi}^{ij},L^{-1}\hat{\pi}^{ij}\rangle \langle\hat{\pi}^{ij},
L^{-1}\hat{\pi}^{ij}\rangle}{\langle\hat{\pi}^{ij},\hat{\pi}^{ij}\rangle}
=\frac{\eta^\mathrm{exact}}{P}.
\end{align}
Eq.~\eqref{eq:setting2} is also verified in an analogous way.

%%%%%%%%%%%%%%%%%%%%%%%%%%%%%%%%%%%%%%%%%%%%%%
\subsection{Convergence properties of the numerical results}
\label{sec:sec5-3}

\begin{figure}[t]
 \begin{minipage}{0.45\hsize}
 \begin{center}
 \includegraphics[width=6.5cm]{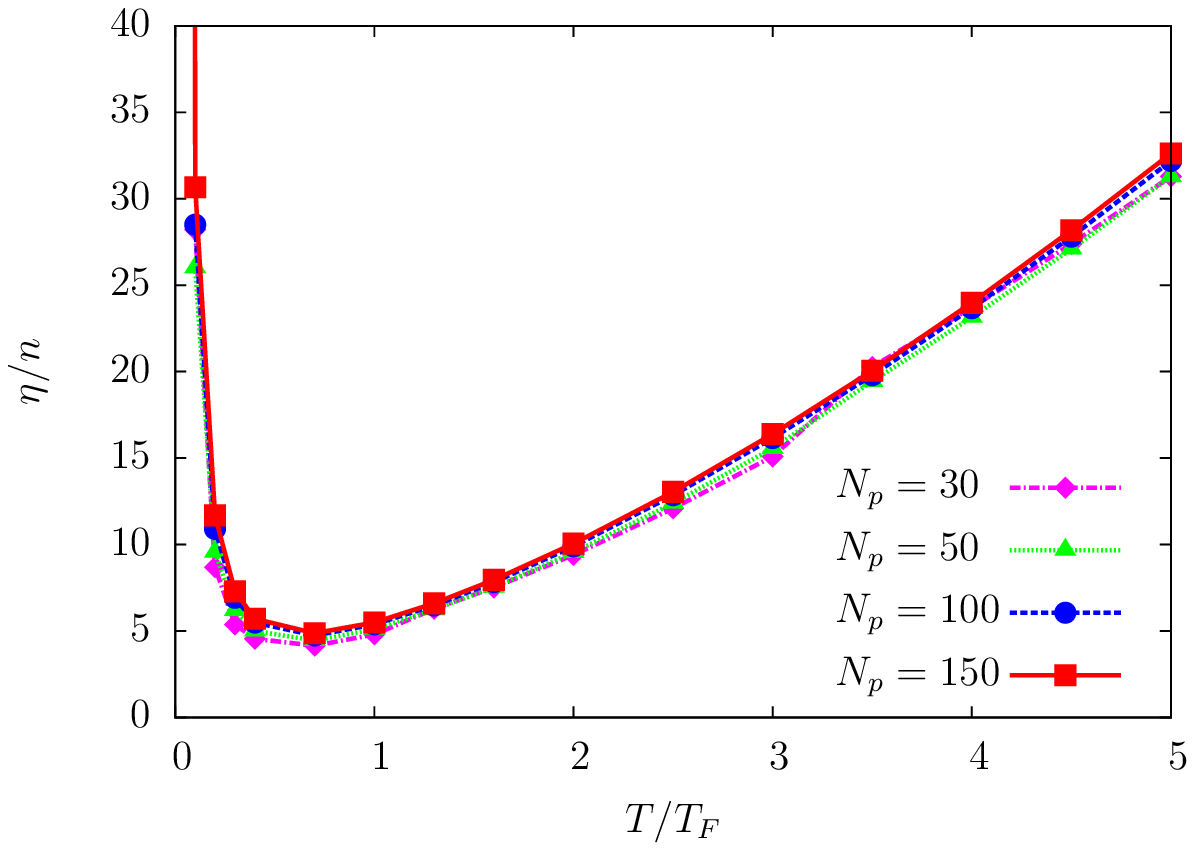}
 \end{center}
 \end{minipage}
 \begin{minipage}{0.45\hsize}
 \begin{center}
 \includegraphics[width=6.5cm]{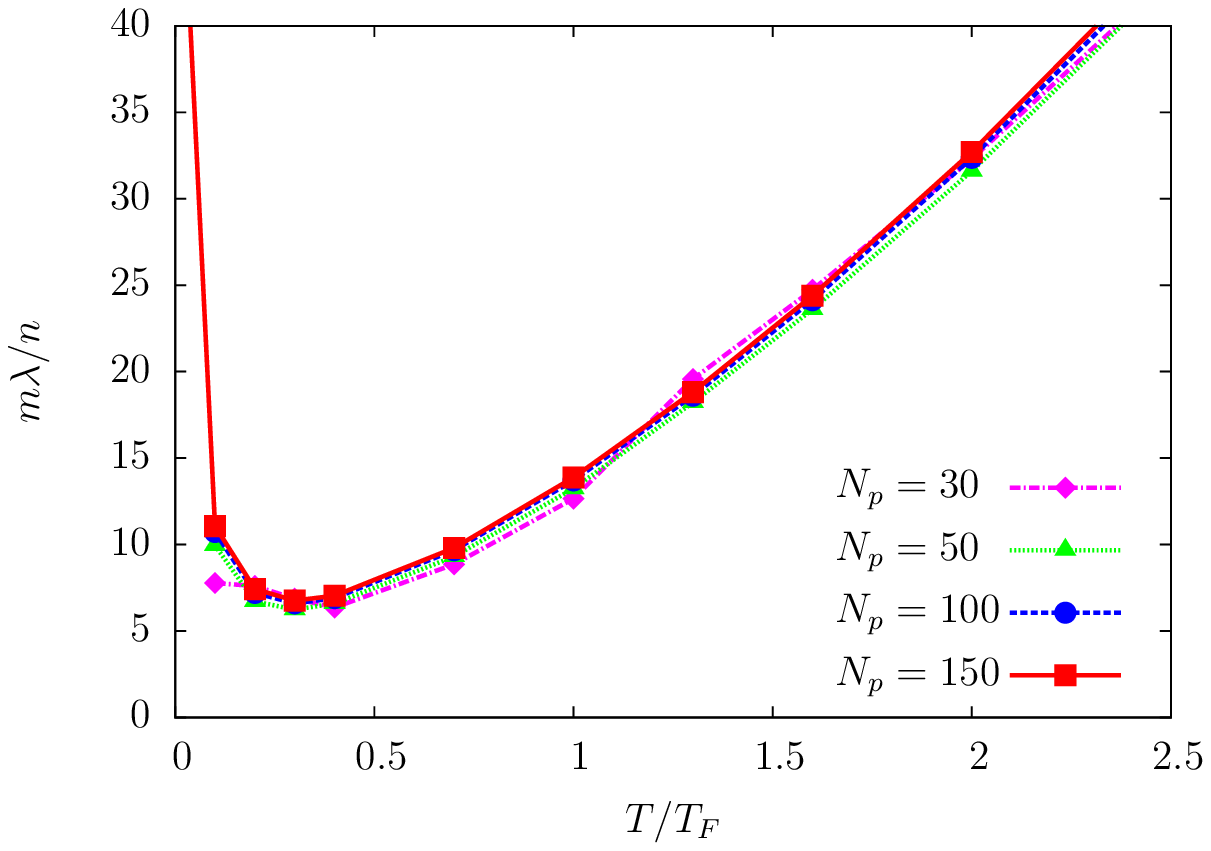}
 \end{center}
 \end{minipage}
 \\
 \begin{minipage}{0.45\hsize}
 \begin{center}
 \includegraphics[width=6.5cm]{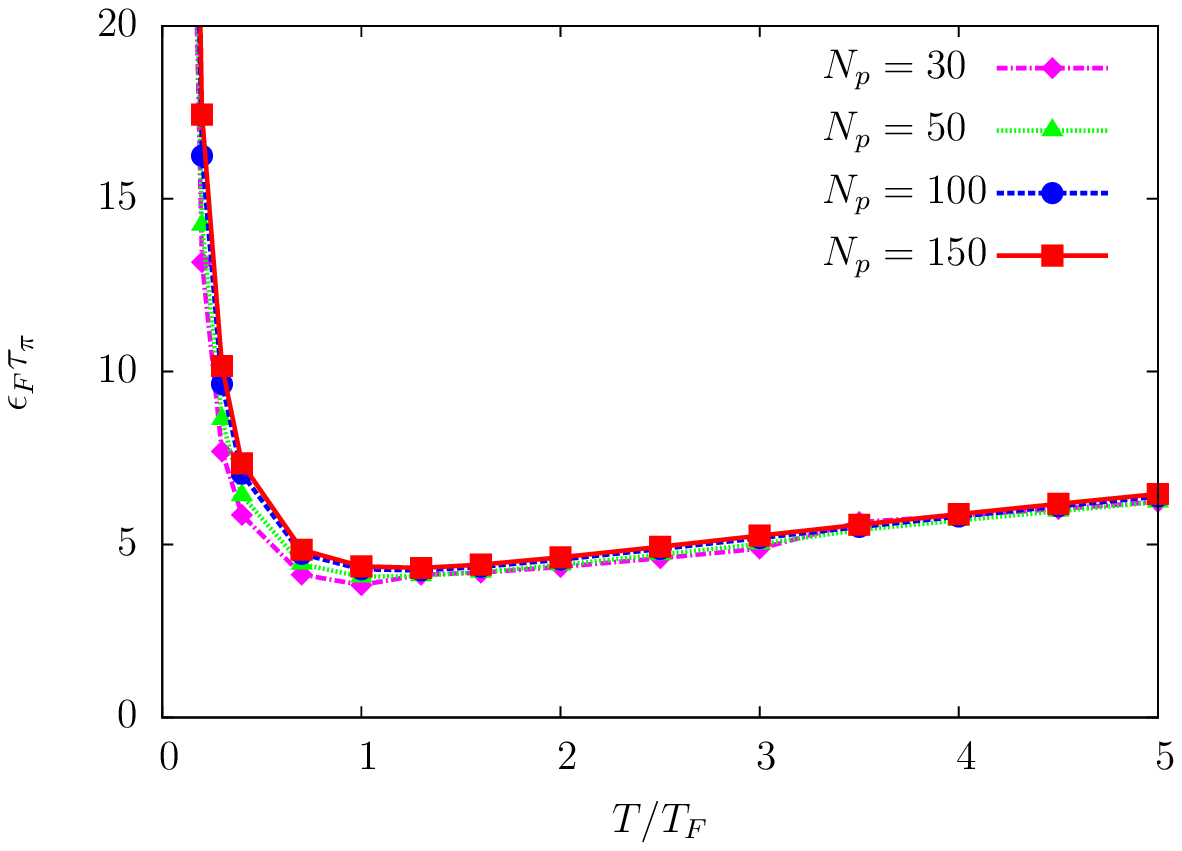}
 \end{center}
 \end{minipage}
 \begin{minipage}{0.45\hsize}
 \begin{center}
 \includegraphics[width=6.5cm]{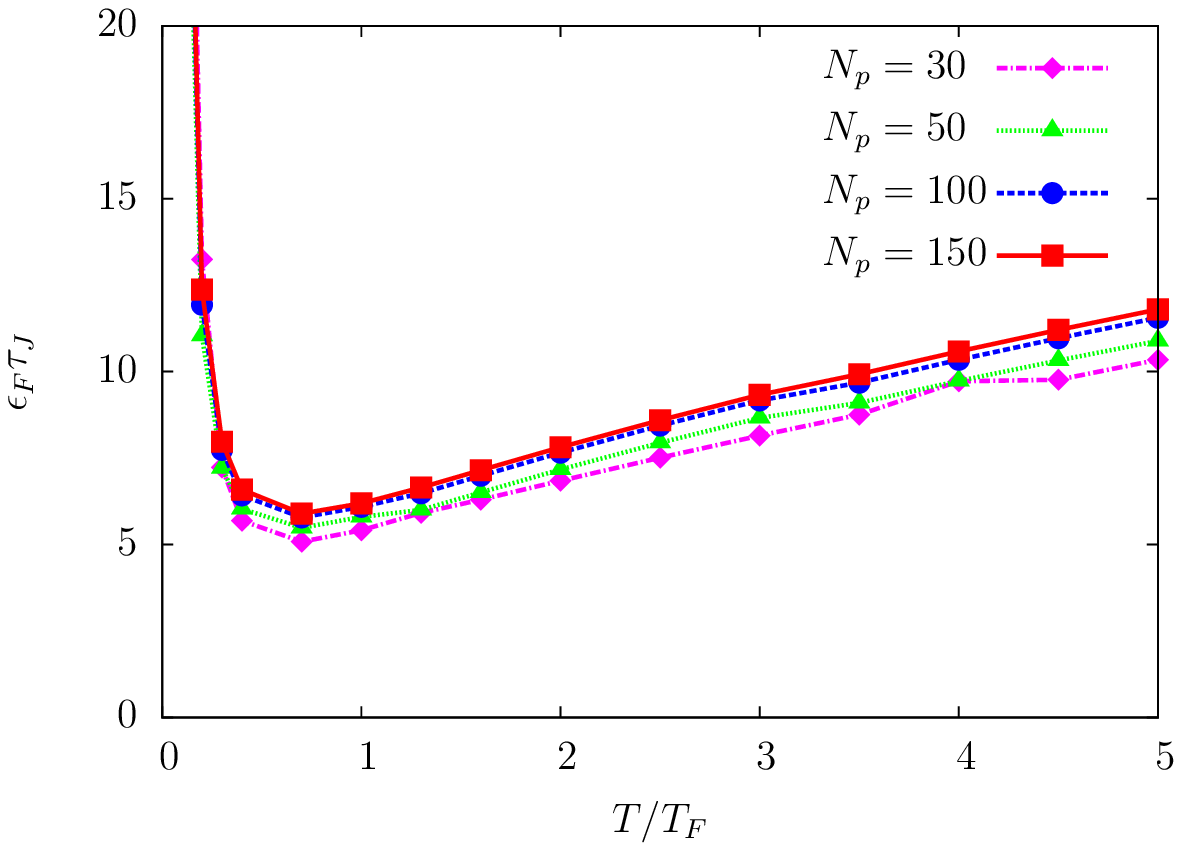}
 \end{center}
 \end{minipage}
 \caption{Temperature dependence of the shear viscosity (left top panel), heat conductivity (right top panel) and viscous relaxation times of the stress tensor (left bottom panel) and heat flow (right bottom panel) at the unitarity. Quantum Fermi statistics is taken into account in these figures. $N_p$ is a number of meshes for the discretized momentum space and the purple rhombus, green triangle, the blue circle and red square indicate $N_p=30,\ 50,\ 100,\ 150$, respectively.}
 \label{fig:T_conv}
\end{figure}

We show that the numerical results of shear viscosity, heat conductivity,
 and viscous relaxation times of the stress tensor and heat flow, 
%which are shown in Sec.~\ref{sec:sec4}, 
are well convergent. In Fig.~\ref{fig:T_conv}, 
$N_p$ is a number of meshes for the discretized momentum space. 
We have confirmed that the temperature dependences of all the quantities shown here are well convergent 
with $N_p=150$. 
In this confirmation, momentum upper cutoff are taken as large enough 
for the numerical results not to depend on them. 
We have also checked their scattering dependence and they are well convergent as well.

%%%%%%%%%%%%%%%%%%%%%%%%%%%%%%%%%%%%%%%%%%%%%%%%%%%%%%%%%%%%%%%%%%%%%

%%%%%%%%%%%%%%%%%%%%%%%%%%%%%%%%%%%%%%%%%%%%%%%%%%%%%%%%%%%%%%%%%%%%%
%\setcounter{equation}{0}
\section{
  Conclusion
}
\label{sec:sec6}

In this paper, we have derived the second-order hydrodynamic equation for non-relativistic systems with the microscopic expressions of all the transport coefficients including the viscous relaxation times by applying the RG method:
It is notable that the shear viscosity and heat conductivity have the same expressions as those by the Chapman-Enskog method, and the viscous relaxation times take new but natural forms.
Though the inclusion of the quantum statistical effects do not change the form of the hydrodynamic equation, it makes the microscopic expressions of the transport coefficients different, which gives the remarkable differences in the value of the transport coefficients. 

By using the transport coefficients that we have obtained, we have calculated in a full numerical way the shear viscosity, heat conductivity and viscous relaxation times of the stress tensor and heat flow.
Any approximation has not been used in the numerical evaluation of these quantities in the present work, which has given the exact values based on the Boltzmann equation, and the numerical convergence is readily confirmed.
We have found that the Fermi statistics makes significant contributions to the first-order transport coefficients and viscous relaxation times at low temperature or small scattering length due to the Pauli blocking in the rigid Fermi sphere, which suppresses the collision rate and results in highly viscous systems.
Furthermore, by using the numerical results we have examined the reliability of the relations $\tau_\pi=\eta/P$ and $\tau_J = 12mT\lambda/(7Q-75P^2/n)$, which are derived with recourse to the RTA and used rather extensively. The resulting data have shown that the ratio of the shear viscosity and pressure well agrees with the viscous relaxation time. This agreement is consistent with Ref.~\cite{Schafer:2014} and suggest the reliability of the RG method. Furthermore, the results encourage us to use the relation $\tau_\pi=\eta/P$, which greatly simplifies the evaluation of the viscous relaxation time of the stress tensor. On the other hand, the latter relation for the viscous relaxation time does not appear to be reliable. Thus, we should use the value evaluated by Eqs.~\eqref{eq:RT1} and \eqref{eq:RT2} instead of Eq.~\eqref{eq:transport_rel} for the viscous relaxation time of the heat conductivity, and also need to investigate the time evolution of fluids with these transport coefficients inserted in order to examine how quantitatively significant the difference of $\tau_J$ evaluated with and without the RTA is.

The numerical evaluation of the other transport coefficients are left as a future work.
Then, we may apply the obtained second-order hydrodynamic equation to the analysis of the time evolution of the ultracold atomic gases.
Furthermore, we should take account of systems with phase transitions. In particular, the behavior of the transport coefficients near the critical region and investigation of how they affect on the time-evolution of fluids are interesting.

%%%%%%%%%%%%%%%%%%%%%%%%%%%%%%%%%%%%%%%%%%%%%%%%%%%%%%%%%%%%%%%%%%%%%

\section*{Acknowledgment}

Y.K. is supported by the Grants-in-Aid for JSPS fellows (No.15J01626). 
T.K. was partially supported by a
Grant-in-Aid for Scientific Research from the Ministry of Education,
Culture, Sports, Science and Technology (MEXT) of Japan
(Nos. 20540265 and 23340067),
by the Yukawa International Program for Quark-Hadron Sciences.

\appendix

%%%%%%%%%%%%%%%%%%%%%%%%%%%%%%%%%%%%%%%%%%%%%%%%%%%%%%%%%%%%%%%%%%%%%
\setcounter{equation}{0}
\section{
  Determination of the initial condition of the first-order perturbative equation
}
\label{sec:app1}

In this appendix, we present a detailed calculation of Eq.~\eqref{eq:LQF}.
$F_{0p}$ given in Eq.~\eqref{eq:F} is calculated as
\begin{align}
 F_{0p}
 &= -(f^{\mathrm{eq}}_p\bar{f}^{\mathrm{eq}}_p)^{-1}\left(\bs{v}\cdot\bs{\nabla}+\bs{F}\cdot\bs{\nabla}_p\right) \tilde{f}^{(i)}_p
 \nonumber\\
 &=-v^i\frac{\nabla^iT}{T^2}\left(\frac{|\bs{\delta p}|^2}{2m}-\mu_{TF}\right) - \frac{1}{T}v^i\delta p^j\nabla^i u^j 
 -  v^i\frac{\nabla^i\mu_{TF}}{T} + F^i\frac{\delta p^i}{mT}.
\end{align}
Then, we can calculate the projection of $F_0$ onto the Q$_0$-space as
\begin{align}
 [Q_0 F_0]_p &= [F_0-P_0 F_0]_p
 = F_{0p} 
 - \sum_{\alpha=0}^4\frac{\varphi^\alpha_{0p}}{c^\alpha}\langle\varphi_0^\alpha,F_0\rangle
 = - \frac{\sigma^{ij}}{T}\hat{\pi}^{ij}_p -\frac{\nabla^i T}{T^2}\hat{J}^i_p,
\end{align}
where we have used the definitions of the projection operators \eqref{eq:Q0} in the first equality and \eqref{eq:P0} in the second equality.
Then, we arrive at Eq.~\eqref{eq:LQF} ,
\begin{align}
 [L^{-1}Q_0 F_0]_p
 = - \frac{\sigma^{ij}}{T}[L^{-1}\hat{\pi}^{ij}]_p - \frac{\nabla^i T}{T^2}[L^{-1}\hat{J}^i]_p.
\end{align}

%%%%%%%%%%%%%%%%%%%%%%%%%%%%%%%%%%%%%%%%%%%%%%%%%%%%%%%%%%%%%%%%%%%%%

%%%%%%%%%%%%%%%%%%%%%%%%%%%%%%%%%%%%%%%%%%%%%%%%%%%%%%%%%%%%%%%%%%%%%
\setcounter{equation}{0}
\section{
  Detailed derivation of relaxation equation
}
\label{sec:app2}

We show a detailed derivation of the relaxation equation and give the explicit expressions of all the coefficients appearing in it. To this end, we reduce Eq.~\eqref{eq:averaging1} into the relaxation equation.
Using the vector notations
\begin{align}
 \hat{\psi}^\alpha_p &= \{\hat{\pi}^{ij}_p, \hat{J}^i_p\},
 \\
 \hat{\chi}^\alpha_p &= \left\{ \frac{\hat{\pi}^{ij}_p}{2T\eta}, \frac{\hat{J}^i_p}{T^2\lambda}\right\},
 \\
 \psi^\alpha &= \{\pi^{ij}, J^i\},
 \\
 X^\alpha &= \left\{ 2\eta\sigma^{ij}, \lambda\nabla^i T\right\},
\end{align} 
with which we can write as $\Psi=L^{-1}\hat{\chi}^\alpha \psi^\alpha$ and $Q_0F_0=-\hat{\chi}^\alpha_1 X^\alpha$.
Equation~\eqref{eq:averaging1} can be converted into the following form
\begin{align}
 \label{eq:rel-eq-2}
 &\epsilon\langle L^{-1}\hat{\psi}^\alpha, \hat{\chi}^\beta \rangle \psi^\beta
 \nonumber\\
 &=\epsilon\langle L^{-1}\hat{\psi}^\alpha, \hat{\chi}^\beta \rangle X^\beta
 +\langle L^{-1}\hat{\psi}^\alpha, L^{-1}\hat{\chi}^\beta \rangle\frac{\mathrm{D}}{\mathrm{D}t}\psi^\beta
 +\epsilon\langle L^{-1}\hat{\psi}^\alpha, \delta K^i L^{-1}\hat{\chi}^\beta \rangle\nabla^i\psi^\beta
 \nonumber\\
 &+\epsilon\Big< L^{-1}\hat{\psi}^\alpha, f^\mathrm{eq}\bar{f}^\mathrm{eq}
 \left(\frac{\mathrm{D}}{\mathrm{D}t} + \epsilon\bs{\delta K}\cdot \bs{\nabla} + \epsilon\bs{F}\cdot \bs{\nabla}_K\right)f^\mathrm{eq}\bar{f}^\mathrm{eq}L^{-1}\hat{\chi}^\beta \Big>\psi^\beta
 \nonumber\\
 &-\epsilon^2\frac{1}{2}\langle L^{-1}\hat{\psi}^\alpha, B[L^{-1}\hat{\chi}^\beta_1,L^{-1}\hat{\chi}^\gamma] \rangle \psi^\beta \psi^\gamma,
\end{align}
where we have defined $[\delta K^i]_p\equiv \delta p^i/m =\delta v^i$ and $[\nabla_K^i]_p\equiv \nabla_p^i$.

The coefficients of the first, third, and fourth terms in the right-hand side of Eq.~\eqref{eq:rel-eq-2} can be written as
\begin{align}
 &\big<L^{-1}\hat{\psi}^\alpha,\hat{\psi}^\beta\big>=
 \begin{pmatrix}
  -2T\eta\Delta^{ijkl}& 0 \\
  0 & -T^2\lambda\Delta^{ij}
 \end{pmatrix},
 \\
 &\big<L^{-1}\hat{\psi}^\alpha,L^{-1}\hat{\psi}^\beta\big>=
 \begin{pmatrix}
  2T\eta\tau_{\pi}\Delta^{ijkl} & 0 \\
  0 & T^2\lambda\tau_{J}\Delta^{ij} \\
 \end{pmatrix},
 \\
 &\big<L^{-1}\hat{\psi}^\alpha,\delta K^m L^{-1}\hat{\psi}^\beta\big>=
 \begin{pmatrix}
  0 & T^2\lambda\ell_{\pi J}\Delta^{ijkl} \\
  2T\eta\ell_{J\pi}\Delta^{ijkl} & 0  \\
 \end{pmatrix},
\end{align}
where transport coefficients introduced here are defined as follows:
\begin{align}
 &\tau_{\pi}\equiv \frac{1}{10T\eta}\big<\hat{\pi}^{ij},L^{-2}\hat{\pi}^{ij}\big>,
 \\
 &\tau_{J}\equiv \frac{1}{3T^2\lambda}\big<\hat{J}^{i},L^{-2}\hat{J}^{i}\big>,
\end{align}
which are viscous relaxation times for the stress tensor and heat flow, respectively, and
\begin{align}
 &\ell_{\pi J}\equiv \frac{1}{5T^2\lambda}
 \big<L^{-1}\hat{\pi}^{ij},\delta K^{i}L^{-1}\hat{J}^{j}\big>,
 \\
 &\ell_{J\pi}\equiv \frac{1}{10T\eta}\big<L^{-1}\hat{J}^{i},\delta K^{j}L^{-1}\hat{\pi}^{ij}\big>,
\end{align}
which are so called viscous relaxation lengths.

Then, let us rewrite the last term in the right-hand side of Eq.~\eqref{eq:rel-eq-2} as
\begin{align}
 &\frac{\epsilon^2}{2}\big<L^{-1}\hat{\pi}^{ij},B[L^{-1}\hat{\chi}^\beta][L^{-1}\hat{\chi}^\gamma]\big>\psi^\beta \psi^\gamma
 =b_{\pi\pi\pi}\pi^{m\langle k}\pi^{l\rangle m} +b_{\pi JJ}J^{\langle k} J^{l\rangle},
 \\
 &\frac{\epsilon^2}{2}\big<L^{-1}\hat{J}^i,B[L^{-1}\hat{\chi}^\beta][L^{-1}\hat{\chi}^\gamma]\big>\psi^\beta \psi^\gamma
 =b_{J\pi J}\pi^{ij} J^j,
\end{align}
where the transport coefficients are defined by
\begin{align}
 &b_{\pi\pi\pi} \equiv \frac{3}{70T^2\eta^2}
 \big<L^{-1}\hat{\pi}^{ij},
 B[L^{-1}\hat{\pi}^{ik}][L^{-1}\hat{\pi}^{jk}]\big>,
 \\
 &b_{\pi JJ} \equiv \frac{1}{10T^4\lambda^2}
 \big<L^{-1}\hat{\pi}^{ij},B[L^{-1}\hat{J}^i][L^{-1}\hat{J}^j]\big>,
 \\
 &b_{JJ\pi} \equiv \frac{1}{10T^3\eta\lambda}
 \big<L^{-1}\hat{J}^i,B[L^{-1}\hat{J}^j][L^{-1}\hat{\pi}^{ij}]\big>.
\end{align}

We consider the forth term in the right-hand side of Eq.~\eqref{eq:rel-eq-2}:
\begin{align}
 &\epsilon\Big<L^{-1}\hat{\psi}^\alpha,(f^{\mathrm{eq}}\bar{f}^{\mathrm{eq}})^{-1}
 \left[\frac{\mathrm{D}}{\mathrm{D}t}+\epsilon\boldsymbol{\delta K}\cdot\bs{\nabla}+\epsilon\bs{F}\cdot\bs{\nabla}_K\right]
 f^{\mathrm{eq}}\bar{f}^{\mathrm{eq}}L^{-1}\hat{\chi}^\beta\Big>\hat{\psi}^\beta
 \nonumber\\
 &=\epsilon\Big<L^{-1}\hat{\psi}^\alpha,(f^{\mathrm{eq}}\bar{f}^{\mathrm{eq}})^{-1}
 \frac{\partial}{\partial T}[f^{\mathrm{eq}}\bar{f}^{\mathrm{eq}}L^{-1}\hat{\chi}^\beta]\Big>
 \hat{\psi}^\beta \frac{\mathrm{D}T}{\mathrm{D}t}
 \nonumber\\
 &+\epsilon^2\Big<L^{-1}\hat{\psi}^\alpha,(f^{\mathrm{eq}}\bar{f}^{\mathrm{eq}})^{-1}
 \delta K^a\frac{\partial}{\partial T}[f^{\mathrm{eq}}\bar{f}^{\mathrm{eq}}L^{-1}\hat{\chi}^\beta]\Big>
 \hat{\psi}^\beta \nabla^a T
 \nonumber\\
 &+\epsilon\Big<L^{-1}\hat{\psi}^\alpha,(f^{\mathrm{eq}}\bar{f}^{\mathrm{eq}})^{-1}
 \frac{\partial}{\partial\mu_{TF}}[f^{\mathrm{eq}}\bar{f}^{\mathrm{eq}}L^{-1}\hat{\chi}^\beta\Big>
 \hat{\psi}^\beta \frac{\mathrm{D}\mu_{TF}}{\mathrm{D}t}
 \nonumber\\
 &+\epsilon^2\Big<L^{-1}\hat{\psi}^\alpha,(f^{\mathrm{eq}}\bar{f}^{\mathrm{eq}})^{-1}
 \delta K^a\frac{\partial}{\partial\mu_{TF}}[f^{\mathrm{eq}}f^{\mathrm{eq}}
 L^{-1}\hat{\chi}^\beta]\Big>\hat{\psi}^\beta \nabla^a \mu_{TF}
 \nonumber\\
 &+\epsilon\Big<L^{-1}\hat{\psi}^\alpha,(f^{\mathrm{eq}}\bar{f}^{\mathrm{eq}})^{-1}
 \frac{\partial}{\partial u^b}[f^{\mathrm{eq}}f^{\mathrm{eq}}L^{-1}\hat{\chi}^\beta]\Big>
 \hat{\psi}^\beta \frac{\mathrm{D}u^b}{\mathrm{D}t}
 \nonumber\\
 &+\epsilon^2\Big<L^{-1}\hat{\psi}^\alpha,(f^{\mathrm{eq}}\bar{f}^{\mathrm{eq}})^{-1}
 \delta K^a\frac{\partial}{\partial u^b}[f^{\mathrm{eq}}f^{\mathrm{eq}}
 L^{-1}\hat{\chi}^\beta]\Big>\hat{\psi}^\beta \nabla^a u^b
 \nonumber\\
 &+\epsilon^2\Big<L^{-1}\hat{\psi}^\alpha,(f^{\mathrm{eq}}\bar{f}^{\mathrm{eq}})^{-1}
 \nabla_K^a[f^{\mathrm{eq}}f^{\mathrm{eq}}L^{-1}\hat{\chi}^\beta]\Big>
 \hat{\psi}^\beta \frac{F^a}{m}.
 \label{2nd-relax4}
\end{align}
The Lagrange derivative of $T$, $\mu_{TF}$, and $u^b$ are rewritten by using the 
balance equation up to the first order with respect to $\varepsilon$, which corresponds to 
the Euler's equation:
\begin{align}
 \frac{\mathrm{D}T}{\mathrm{D}t}
 &=-\epsilon\frac{2T}{3}\bs{\nabla}\cdot\bs{u}
 +\mathcal{O}(\epsilon^2),
 \\
 \label{eq:Euler-mu}
 \frac{\mathrm{D}\mu_{TF}}{\mathrm{D}t}
 &=-\epsilon\frac{2\mu_{TF}}{3}\boldsymbol{\nabla}\cdot\boldsymbol{u}
 +\mathcal{O}(\epsilon^2),
 \\
 \frac{\mathrm{D}u^b}{\mathrm{D}t}
 &=-\epsilon\nabla^i P +\epsilon nF^i
 +\mathcal{O}(\epsilon^2),
\end{align}
where we have used the relation $\mathrm{d}n=(\partial n/\partial T)\mathrm{d}T +(\partial n/\partial\mu_{TF})\mathrm{d}\mu_{TF}$ in the derivation of Eq.~\eqref{eq:Euler-mu}.
Then, Eq.~\eqref{2nd-relax4} takes the following forms
\begin{align}
 &\epsilon\Big<L^{-1}\hat{\pi}^{ij},(f^{\mathrm{eq}}\bar{f}^{\mathrm{eq}})^{-1}
 \left[\frac{\mathrm{D}}{\mathrm{D}t}+\epsilon\boldsymbol{\delta K}\cdot\bs{\nabla}+\epsilon\bs{F}\cdot\bs{\nabla}_K\right]
 f^{\mathrm{eq}}\bar{f}^{\mathrm{eq}}\hat{\chi}^\beta\Big>\psi^\beta
 \nonumber\\
 &=-\epsilon^2\left(\kappa_{\pi\pi}^{(1)}\pi^{ij}\boldsymbol{\nabla}\cdot\boldsymbol{u}
 +\kappa_{\pi\pi}^{(2)}\pi^{k\langle i}\sigma^{j\rangle k}
 +\kappa_{\pi\pi}^{(3)}\pi^{k\langle i}\omega^{j\rangle k}
 +\kappa_{\pi J}^{(1)}J^{\langle i}\nabla^{j\rangle}n
 +\kappa_{\pi J}^{(2)}J^{\langle i}\nabla^{j\rangle}P
 +\kappa_{\pi J}^{(3)}J^{\langle i}F^{j\rangle}\right),
 \\[10pt]
 &\epsilon\Big<L^{-1}\hat{J}^k,(f^{\mathrm{eq}}\bar{f}^{\mathrm{eq}})^{-1}
 \left[\frac{\mathrm{D}}{\mathrm{D}t}+\epsilon\bs{\delta K}\cdot\bs{\nabla}+\epsilon\bs{F}\cdot\bs{\nabla}_K\right]
 f^{\mathrm{eq}}\bar{f}^{\mathrm{eq}}L^{-1}\hat{\chi}^\beta\Big>\psi^\beta
 \nonumber\\
 &=-\epsilon^2\left(\kappa_{J\pi}^{(1)}\pi^{ij}\nabla^j n
 +\kappa_{J\pi}^{(2)}\pi^{ij}\nabla^j P
 +\kappa_{J\pi}^{(3)}\pi^{ij}F^j
 +\kappa_{JJ}^{(1)}J^i\boldsymbol{\nabla}\cdot\boldsymbol{u}
 +\kappa_{J\pi}^{(2)}J^j\sigma^{ij}
 +\kappa_{J\pi}^{(3)}J^j\omega^{ij}\right),
\end{align}
where we have used $\nabla^i u^j=\sigma^{ij}+\omega^{ij}+\delta^{ij}\boldsymbol{\nabla}\cdot\boldsymbol{u}/3$ with the vorticity $\omega^{ij}\equiv(\nabla^i u^j-\nabla^j u^i)/2$, and the transport coefficients are defined as follows:
\begin{align}
 \kappa_{\pi\pi}^{(1)}
 &\equiv -\frac{1}{10}
 \Big<L^{-1}\hat{\pi}^{ij},(f^{\mathrm{eq}}\bar{f}^{\mathrm{eq}})^{-1}
 \left[-\frac{2T}{3}\frac{\partial}{\partial T}
 -\frac{2\mu_{TF}}{3}\frac{\partial}{\partial \mu_{TF}}
 +\frac{1}{3}\delta K^a \frac{\partial}{\partial u^a}\right]
 \frac{f^{\mathrm{eq}}\bar{f}^{\mathrm{eq}}L^{-1}\hat{\pi}^{ij}}{T\eta}\Big>,
 \\
 \kappa_{\pi\pi}^{(2)}
 &\equiv -\frac{6}{35}\Delta^{kjab}
 \Big<L^{-1}\hat{\pi}^{ij},(f^{\mathrm{eq}}\bar{f}^{\mathrm{eq}})^{-1}
 \delta K^a\frac{\partial}{\partial u^b}
 \frac{f^{\mathrm{eq}}\bar{f}^{\mathrm{eq}}L^{-1}\hat{\pi}^{ki}}{T\eta}\Big>,
 \\
 \label{eq:kappapipi3}
 \kappa_{\pi\pi}^{(3)}
 &\equiv -\frac{2}{15}\Omega^{kjab}
 \Big<L^{-1}\hat{\pi}^{ij},(f^{\mathrm{eq}}\bar{f}^{\mathrm{eq}})^{-1}
 \delta K^a\frac{\partial}{\partial u^b}
 \frac{f^{\mathrm{eq}}\bar{f}^{\mathrm{eq}}L^{-1}\hat{\pi}^{ki}}{T\eta}\Big>
 = -2\tau_\pi,
 \\
 \kappa_{\pi J}^{(1)}
 &\equiv -\frac{1}{5}
 \Big<L^{-1}\hat{\pi}^{ij},(f^{\mathrm{eq}}\bar{f}^{\mathrm{eq}})^{-1}\delta K^i
 \frac{2T^2}{3nT-2Ah_{TF}}\left[\frac{\partial}{\partial T}-s\frac{\partial}{\partial\mu_{TF}}\right]
 \frac{f^{\mathrm{eq}}\bar{f}^{\mathrm{eq}}L^{-1}\hat{J}^j}{T^2\lambda}\Big>,
 \\
 \kappa_{\pi J}^{(2)}
 &\equiv -\frac{1}{5}
 \Big<L^{-1}\hat{\pi}^{ij},(f^{\mathrm{eq}}\bar{f}^{\mathrm{eq}})^{-1}
 \nonumber\\
 &\times\left[-\delta K^i\frac{2AT/n}{3nT-2Ah_{TF}}\frac{\partial}{\partial T} + \delta K^i\frac{1}{n}\frac{\partial}{\partial\mu_{TF}} - \frac{\partial}{\partial u^i}\right]
 \frac{f^{\mathrm{eq}}\bar{f}^{\mathrm{eq}}L^{-1}\hat{J}^j}{T^2\lambda}\Big>,
 \\
 \kappa_{\pi J}^{(3)}
 &\equiv -\frac{1}{5}
 \Big<L^{-1}\hat{\pi}^{ij},(f^{\mathrm{eq}}\bar{f}^{\mathrm{eq}})^{-1}
 \left[n\frac{\partial}{\partial u^i} + \nabla_K^i\right]
 \frac{f^{\mathrm{eq}}\bar{f}^{\mathrm{eq}}L^{-1}\hat{J}^j}{T^2\lambda}\Big>,
 \\
 \kappa_{J\pi}^{(1)}
 &\equiv -\frac{1}{10}
 \Big<L^{-1}\hat{J}^i,(f^{\mathrm{eq}}\bar{f}^{\mathrm{eq}})^{-1}\delta K^j
 \frac{2T^2}{3nT-2Ah_{TF}}\left[\frac{\partial}{\partial T}-s\frac{\partial}{\partial\mu_{TF}}\right]
 \frac{f^{\mathrm{eq}}\bar{f}^{\mathrm{eq}}L^{-1}\hat{\pi}^{ij}}{T\eta}\Big>,
 \\
 \kappa_{J\pi}^{(2)}
 &\equiv -\frac{1}{10}
 \Big<L^{-1}\hat{J}^i,(f^{\mathrm{eq}}\bar{f}^{\mathrm{eq}})^{-1}
 \nonumber\\
 &\times\left[-\delta K^j\frac{2AT/n}{3nT-2Ah_{TF}}\frac{\partial}{\partial T}+\delta K^j\frac{1}{n}\frac{\partial}{\partial\mu_{TF}} - \frac{\partial}{\partial u^j}\right]
 \frac{f^{\mathrm{eq}}\bar{f}^{\mathrm{eq}}L^{-1}\hat{\pi}^{ij}}{T\eta}\Big>,
 \\
 \kappa_{J\pi}^{(3)}
 &\equiv -\frac{1}{10T\eta}
 \Big<L^{-1}\hat{J}^i,(f^{\mathrm{eq}}\bar{f}^{\mathrm{eq}})^{-1}
 \left[n\frac{\partial}{\partial u^j} + \nabla_K^j\right]
 \frac{f^{\mathrm{eq}}\bar{f}^{\mathrm{eq}}L^{-1}\hat{\pi}^{ij}}{T\eta}\Big>,
 \\
 \kappa_{JJ}^{(1)}
 &\equiv -\frac{1}{3}
 \Big<L^{-1}\hat{J}^i,(f^{\mathrm{eq}}\bar{f}^{\mathrm{eq}})^{-1}
 \left[-\frac{2T}{3}\frac{\partial}{\partial T}
 -\frac{2\mu_{TF}}{3}\frac{\partial}{\partial \mu_{TF}}
 +\frac{1}{3}\delta K^a \frac{\partial}{\partial u^a}\right]
 \frac{f^{\mathrm{eq}}\bar{f}^{\mathrm{eq}}L^{-1}\hat{J}^i}{T^2\lambda}\Big>,
 \\
 \kappa_{JJ}^{(2)}
 &\equiv -\frac{1}{5}\Delta^{ijkl}
 \Big<L^{-1}\hat{J}^i,(f^{\mathrm{eq}}\bar{f}^{\mathrm{eq}})^{-1}
 \delta K^k\frac{\partial}{\partial u^l}
 \frac{f^{\mathrm{eq}}\bar{f}^{\mathrm{eq}}L^{-1}\hat{J}^j}{T^2\lambda}\Big>,
 \\
 \label{eq:kappaJJ3}
 \kappa_{JJ}^{(3)}
 &\equiv -\frac{1}{3}\Omega^{ijkl}
 \Big<L^{-1}\hat{J}^i,(f^{\mathrm{eq}}\bar{f}^{\mathrm{eq}})^{-1}
 \delta K^k\frac{\partial}{\partial u^l}
 \frac{f^{\mathrm{eq}}\bar{f}^{\mathrm{eq}}L^{-1}\hat{J}^j}{T^2\lambda}\Big>
 = \tau_J,
\end{align}
where $\Omega^{ijkl}\equiv(\delta^{ik}\delta^{jl}-\delta^{il}\delta^{jk})/2$ is an antisymmetric projection operator, and  $A$ is defined by $A\equiv T\partial n/\partial \mu_\mathrm{TF}$.

Here, we show that 
$\kappa_{\pi\pi}^{(3)}=-2\tau_\pi$ by analytically evaluating the inner product of Eq.~\eqref{eq:kappapipi3}.
Without loss of generality, $L^{-1}\hat{\pi}^{\mu\nu}$ may be written as
\begin{align}
 L^{-1}\hat{\pi}^{ij} = C(|\bs{\delta K}|)\Delta^{ijkl}\delta K^k \delta K^l.
\end{align}
We do not need the specific form of $C(|\bs{\delta K}|)$ in this computation.
Then $\kappa_{\pi\pi}^{(3)}$ can be written as
\begin{align}
 \label{eq:kappa3-2}
 \kappa_{\pi\pi}^{(3)}
 = -\frac{2}{15}\Omega^{kjab}
 \Big<C(|\bs{\delta K}|)\Delta^{ijcd}\delta K^c \delta K^d,(f^{\mathrm{eq}}\bar{f}^{\mathrm{eq}})^{-1}
 \delta K^a\frac{\partial}{\partial u^b}
 \frac{f^{\mathrm{eq}}\bar{f}^{\mathrm{eq}}C(|\bs{\delta K}|)\Delta^{kief}\delta K^e \delta K^f}{T\eta}\Big>.
\end{align}
Here we write down useful formulae for further conversion:
\begin{align}
 \Omega^{kjab}\delta K^a\frac{\partial}{\partial u^b} |\bs{\delta K}|
 &=\Omega^{kjab}\delta K^a\frac{-\delta K^b}{|\bs{\delta K}|}
 =0,
 \\
 \Omega^{kjab}
 \Delta^{ijcd}\delta K^c \delta K^d \delta K^a\Delta^{kief}[-\delta^{be} \delta K^f - \delta^{bf} \delta K^e]
 &= \frac{3}{2}\Delta^{ijkl}\delta K^i \delta K^j \delta K^k \delta K^l
 \nonumber\\
 &=\frac{3}{2}\Delta^{ijkl}\delta K^k \delta K^l \Delta^{ijab}\delta K^a \delta K^b.
\end{align}
By using these formulae, Eq.~\eqref{eq:kappa3-2} is calculated to be
\begin{align}
 \label{eq:kappa3-3}
 \kappa_{\pi\pi}^{(3)}
 &= -\frac{2}{15}\Omega^{kjab}
 \Big<C(|\bs{\delta K}|)\Delta^{ijcd}\delta K^c \delta K^d,
 \delta K^a \frac{C(|\bs{\delta K}|)\Delta^{kief}(-\delta^{be} \delta K^f - \delta^{bf} \delta K^e)}{T\eta}\Big>
 \nonumber\\
 &=-\frac{2}{15}\frac{3/2}{2T\eta}\langle \hat{L}^{-1}\hat{\pi}^{\mu\nu}, \hat{L}^{-1} \hat{\pi}_{\mu\nu} \rangle
 \nonumber\\
 &=-2\tau_\pi.
\end{align}
Similarly we can show that $\kappa_{JJ}^{(3)}=\tau_J$ by evaluating the inner product of Eq.~\eqref{eq:kappaJJ3}.

Combining the formulas derived so far, 
we can rewrite the relaxation equation Eq.~\eqref{eq:rel-eq-2} in the following forms:
\begin{align}
  \epsilon\pi^{ij}
  &= \epsilon2\eta\sigma^{ij}
  - \epsilon^2\tau_\pi \frac{\mathrm{D}}{\mathrm{D}t}\pi^{ij}
  - \epsilon^2\ell_{\pi J}\nabla^{\langle i} J^{j\rangle} 
  \nonumber\\
  &+\epsilon^2 \kappa_{\pi\pi}^{(1)}\pi^{ij}\boldsymbol{\nabla}\cdot\boldsymbol{u}
 +\epsilon^2 \kappa_{\pi\pi}^{(2)}\pi^{k\langle i}\sigma^{j\rangle k}
 -\epsilon^2 2\tau_\pi \pi^{k\langle}\omega^{j\rangle k}
 \nonumber\\
 &+\epsilon^2 \kappa_{\pi J}^{(1)}J^{\langle i}\nabla^{j\rangle}n
 +\epsilon^2 \kappa_{\pi J}^{(2)}J^{\langle i}\nabla^{j\rangle}P
 +\epsilon^2 \kappa_{\pi J}^{(3)}J^{\langle i}F^{j\rangle}
  \nonumber\\
  &+ \epsilon^2 b_{\pi\pi\pi} \pi^{k\langle i} \pi^{j\rangle k}
  + \epsilon^2 b_{\pi JJ} J^{\langle i} J^{j\rangle},
\\[10pt]
  \epsilon J^i
  &= \epsilon\lambda \nabla^i T
  - \epsilon^2\tau_J \frac{\mathrm{D}}{\mathrm{D}t}J^i
  - \epsilon^2\ell_{J\pi}\nabla^j \pi^{ij}
  \nonumber\\
  &+\epsilon^2 \kappa_{J\pi}^{(1)}\pi^{ij}\nabla^j n
 +\epsilon^2 \kappa_{J\pi}^{(2)}\pi^{ij}\nabla^j P
 +\epsilon^2 \kappa_{J\pi}^{(3)}\pi^{ij}F^j
 \nonumber\\
 &+\epsilon^2 \kappa_{JJ}^{(1)}J^i\boldsymbol{\nabla}\cdot\boldsymbol{u}
 +\epsilon^2 \kappa_{JJ}^{(2)}J^j\sigma^{ij}
 +\epsilon^2 \tau_J J^j\omega^{ij}
  \nonumber\\
  &+\epsilon^2 b_{JJ\pi}J^j \pi^{ij}.
\end{align}
Putting back $\epsilon=1$, we arrive at Eqs.~\eqref{eq:relax1} and \eqref{eq:relax2}.

%%%%%%%%%%%%%%%%%%%%%%%%%%%%%%%%%%%%%%%%%%%%%%%%%%%%%%%%%%%%%%%%%%%%%

\end{document}